\documentclass[journal]{IEEEtran}
\ifCLASSINFOpdf
\else
\fi
\usepackage{cite}
\usepackage{url}
\usepackage{hyperref}
\usepackage{graphicx}
\usepackage[small]{caption}
\usepackage{subcaption}
\usepackage{color,soul}
\usepackage{array}
\usepackage{amsmath}
\usepackage{booktabs}
\usepackage{footnote}
\makesavenoteenv{tabular}
\makesavenoteenv{table}
\usepackage{color}
\usepackage{amssymb}
\usepackage[linesnumbered,ruled,vlined]{algorithm2e}
\usepackage[noend]{algpseudocode}

\begin{document}
%
\title{PBMap: A Path Balancing Technology Mapping Algorithm for Single Flux Quantum Logic Circuits}

\author{  Ghasem Pasandi,~\IEEEmembership{Student Member,~IEEE,} and Massoud Pedram,~\IEEEmembership{Fellow,~IEEE}

\thanks{Authors are with the Department of Electrical Engineering, University of Southern California, Los Angeles,
California, USA. e-mails: pasandi@usc.edu, pedram@usc.edu
}

\thanks{Manuscript submitted June 26, 2018.}
}

\markboth{IEEE Transactions on Applied Superconductivity (DOI: 10.1109/TASC.2018.2880343)}
{Pasandi and Pedram: PBMap: A Path Balancing Technology Mapping Algorithm for Rapid Single Flux Quantum Logic Circuits}

%

\maketitle

\begin{abstract}
This paper presents a path balancing technology mapping algorithm, which is a new algorithm for generating a mapping solution for a given Boolean network such that the average logic level difference among fanin gates of each gate in the network is minimized. Path balancing technology mapping is required in dc-biased Single Flux Quantum (SFQ) circuits for guaranteeing the correct operation, and it is beneficial in CMOS circuits to reduce the hazard issues. We present a dynamic programming based algorithm for path balancing technology mapping which generates optimal solutions for dc-biased SFQ (e.g. Rapid SFQ or RSFQ) circuits with tree structure and acts as an effective heuristic for circuits with general Directed Acyclic Graph (DAG) structure. Experimental results show that our path balancing technology mapper reduces the balancing overhead by up to $2.7\times$ and with an average of 21\% compared to the state-of-the-art academic technology mappers.
\end{abstract}

\IEEEpeerreviewmaketitle

\section{Introduction}
\label{Intro:sec}
\IEEEPARstart{P}{ath} balancing technology mapping is a new method of mapping an RTL description such as a Boolean network into a gate-level netlist. For a network generated by the path balancing mapper, the average logic level\footnote{Logic level of a gate $g_i$ in a network $N$ is the length of the longest path (in terms of the gate count) from any primary input of $N$ to $g_i$.} difference among fanin gates of each gate in the mapped netlist is reduced (ideally zero). The path balancing technology mapping is required in Single Flux Quantum (SFQ) logic families including Rapid Single Flux Quantum (RSFQ) \cite{likharev1991rsfq}, energy-efficient SFQ (eSFQ) \cite{volkmann2013experimental}, and Energy-efficient RSFQ (ERSFQ) \cite{kirichenko2011zero} for correct circuit operation. 

SFQ gates with switching delay of $1ps$ and switching energy of $10^{-19}J$ are potential candidates for replacing CMOS gates to achieve high performance and energy efficient systems \cite{volkmann2013experimental}. As an example, a T-Flip-Flop (TFF) with speed of 770GHz is reported in \cite{chen1999rapid}. SFQ circuits are made of Josephson Junctions (JJs), which are superconducting devices working based on the Josephson effect \cite{holmes2013energy}. One of the most popular families of SFQ circuits is RSFQ, which is developed in 1980s \cite{likharev1991rsfq}. 

Due to some key differences between SFQ and CMOS circuits such as gate-level pipelining and fanout limitation in SFQ circuits, existing Computer-Aided Design (CAD) tools for CMOS technology cannot be directly used for SFQ circuits \cite{fourie2018digital, katam2017design}. Therefore, to make use of benefits that SFQ circuits provide in generating high performance and low-power solutions, new design concepts, automation tools, and architectures are needed \cite{holmes2013energy}.
An example difference between SFQ and CMOS gates is that most of SFQ gates (except for confluence buffers, splitters, TFFs and I/O cells) receive a clock signal. This makes the clock distribution network in SFQ more complex than CMOS; the clock network is much bigger in SFQ compared to CMOS and it should be designed more carefully to guarantee delivery of clock signals to all gates with acceptable amounts of jitter and skew \cite{friedman2001clock,shahsavani2017integrated}. 

Another difference between SFQ and CMOS circuits is the requirement of path balancing in SFQ circuits; if there is a difference among logic levels of fanin gates of a gate in an SFQ circuit, path balancing D-Flip-Flops (DFFs) should be inserted into outputs of the fanin gates with smaller logic levels. This is done to guarantee arrival of all input signals of a gate at the same clock period. Otherwise, input pulses which have arrived at earlier clock periods will be consumed, generating wrong output values. For some small circuits, one could add a few asynchronous delay elements (e.g. chain of Josephson Transmission Lines (JTLs)) to make sure that all gates receive their inputs in right clock periods, hence, guaranteeing correct circuit operation. However, this solution cannot be scaled and it is hard to be automated, because it requires information of routed wires (after place and route) during logic synthesis. 
Therefore, path balancing should be considered in logic synthesis (e.g. during the technology mapping phase) of SFQ circuits not only to meet the balancing requirement, but to minimize the path balancing overhead by reducing the number of required path balancing DFFs.
\begin{figure*}[t]
        \centering
        \begin{subfigure}[!t]{0.34\textwidth}
                \centering
                \includegraphics[width=\textwidth]{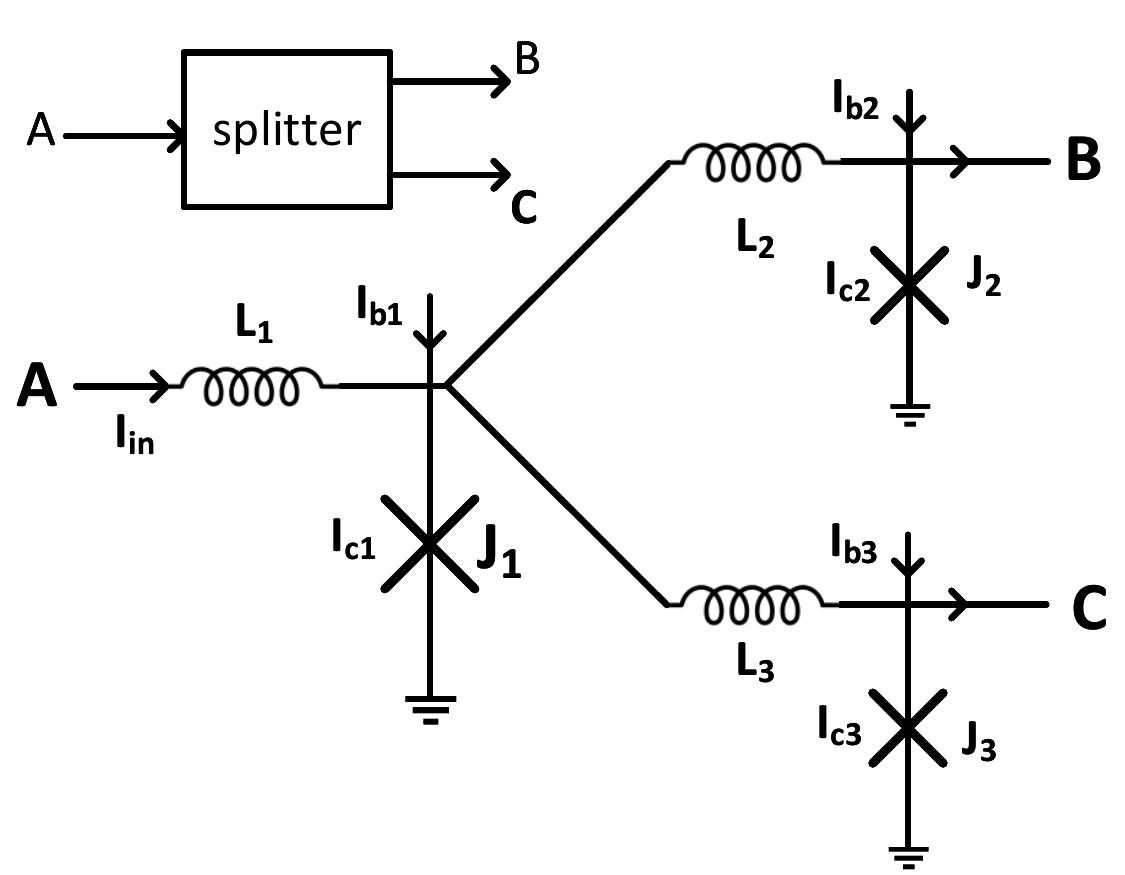}
                \caption{}
                \label{6T_cell}
        \end{subfigure}
        \begin{subfigure}[!t]{0.3\textwidth}
                \centering
                \includegraphics[width=0.95\textwidth]{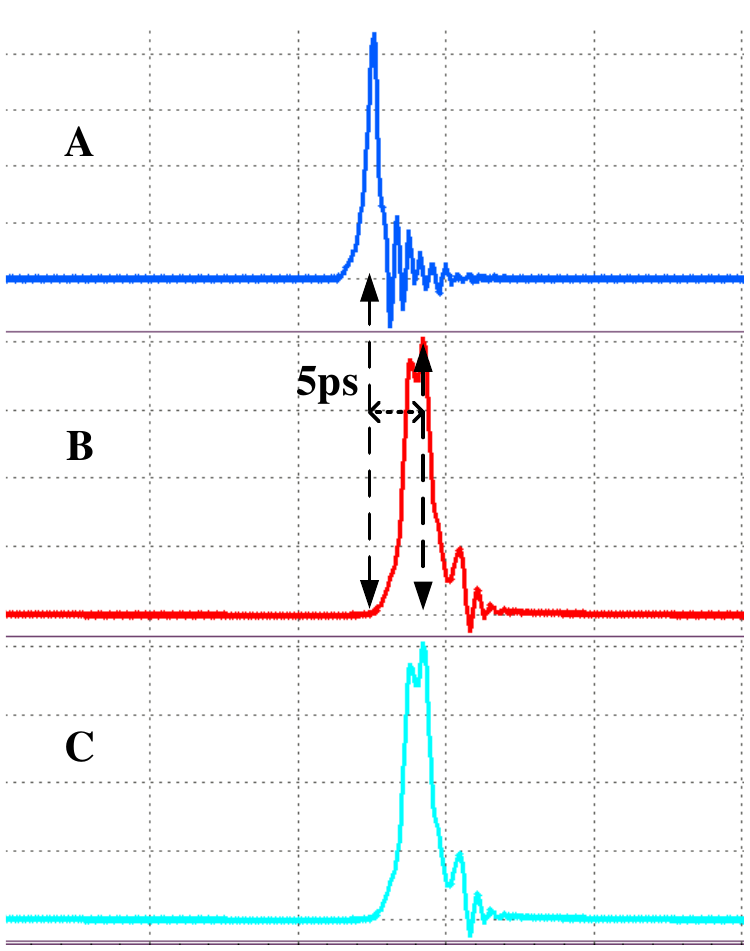}
                \caption{}
                \label{6T_Layout}
        \end{subfigure}
        \begin{subfigure}[!t]{0.35\textwidth}
                \centering
                \includegraphics[width=0.95\textwidth]{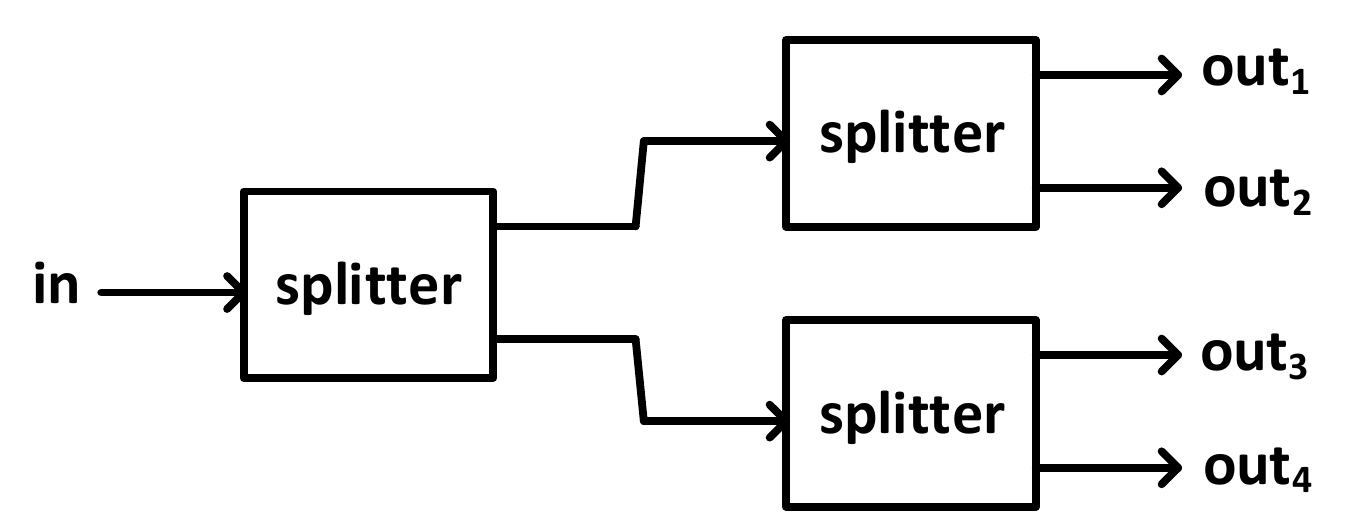}
                \caption{}
                \label{6T_Layout}
        \end{subfigure}
        \begin{subfigure}[!t]{0.37\textwidth}
                \centering
                \includegraphics[width=\textwidth]{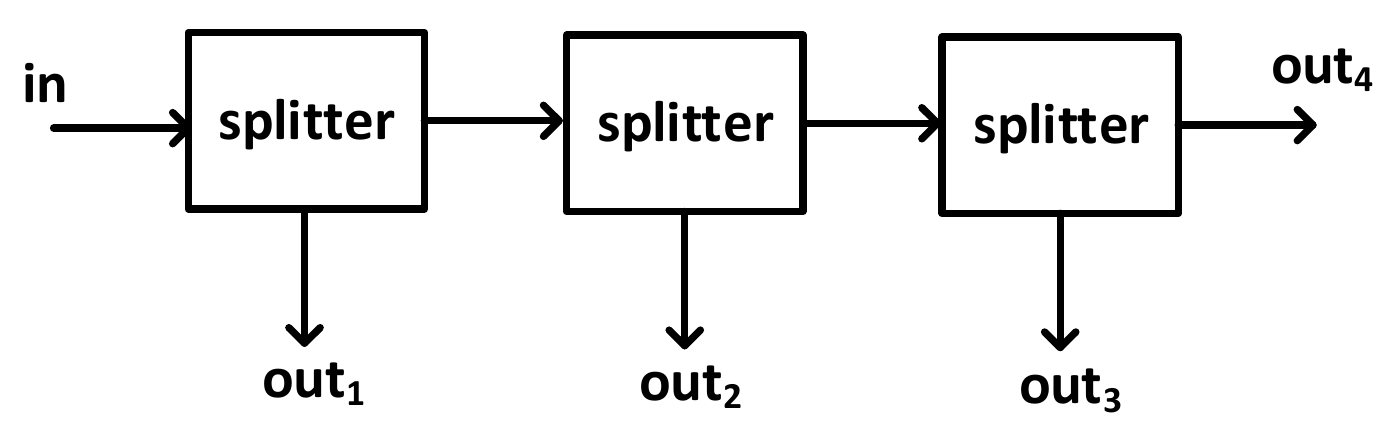}
                \caption{}
                \label{6T_Layout}
        \end{subfigure}
        \caption{(a) Splitter gate in SFQ, (b) waveforms corresponding to the operation of this gate, and a splitter tree providing four fanouts with depth (c) 2, and (d) 3. Delay for generating $out_1$ in \textit{(d)} is lower than \textit{(c)}. Thus, using the structure of \textit{(d)} is better in networks where the critical path goes through $out_1$.}
        \label{Splitter}
\end{figure*}

In this paper, we present PBMap: a path balancing technology mapping algorithm which provides optimal solutions for mapping tree-like dc-biased SFQ logic (including RSFQ, eSFQ, and ERSFQ) circuits. Note that ERSFQ logic was developed to eliminate static power losses of RSFQ by replacing bias resistors with inductors and current-limiting Josephson junctions. Similarly, eSFQ logic, which was also powered by direct current, differed from ERSFQ in the size of the bias current limiting inductor and how the limiting JJs were regulated. So, although there are key differences among RSFQ, ERSFQ, and eSFQ in terms of their biasing network designs, these differences do not affect the proposed mapping algorithm. 

In our algorithm, DFF insertion to achieve path balancing is done to enable gate-level wave-pipelining. In other words, in a circuit generated by our algorithm, length of all paths from any Primary Input (PI) to any Primary Output (PO) will be the same. However, the benefit of using our path balancing algorithm is that it reduces total number of required path balancing DFFs and as a result it reduces total JJ count and total area (Table \ref{exp_table}). 

The rest of this paper is organized as follows: Section \ref{Prior-Motiv:sec} provides some background knowledge on SFQ circuits and logic synthesis, and it summarizes the related work. It also gives a quick overview on the state-of-the-art technology mapping flow. Section \ref{DP_Tech-Map} provides a motivation example for path balancing technology mapping, presents our path balancing technology mapping algorithm, gives its proof of optimality for trees, considers retiming, generalizes the technology mapping algorithm to Directed Acyclic Graphs (DAGs), and finally talks about clock jitter accumulation problem. Section \ref{exper:sec} gives the experimental results, and finally, Section \ref{conc:sec} concludes the manuscript. 

\section{Background}
\label{Prior-Motiv:sec}
\subsection{Background on SFQ Logic Circuits}
\label{BG-RSFQ:sec}
In SFQ logic, a single quanta of magnetic flux ($\Phi_0$ = $h/2e$ = $2.07 mV\times ps$) is used for representation of logic bits. In this representation, presence of a pulse has the meaning of ``logic-1", while absence of a pulse is considered as a ``logic-0". Operation of SFQ logic is based on overdamped Josephson junctions, and hence, it does not experience the problem of hysteretic I-V, which degrades the operation speed of ``1" to ``0" switching.

SFQ logic families are divided into two groups: ac-biased and dc-biased. Adiabatic Quantum Flux Parametron (AQFP) \cite{takeuchi2013adiabatic, takeuchi2014energy} and Reciprocal Quantum Logic (RQL) \cite{herr2011ultra} are examples of ac-biased and RSFQ \cite{likharev1991rsfq} is an example for dc-biased logic family.
The first version of new SFQ logic relied on having ohmic resistors for interconncetion of JJs, hence, is called Resistive Single Flux Quantum logic \cite{likharev1985resistive}. Using this logic, the operation speed of up to 30GHz was reported, which was quite higher than any other digital device with the same complexity at that time \cite{koshelets1987experimental}. Later on, another version was proposed by using JJs instead of ohmic resistors. This version is called Rapid SFQ (RSFQ) \cite{likharev1991rsfq}. It improved the parameter margins of the first version and also increased its operation speed to 300GHz \cite{mukhanov1987ultimate}. In the following, we explain some properties and key circuit/gate level requirements of SFQ circuits.

\subsubsection{Fanout in SFQ}\label{FanoutSubSubSec}In SFQ logic (RSFQ/ERSFQ/eSFQ), if a gate needs to have more than one fanout, a special SFQ gate called \textit{splitter} should be added to the output of this gate. Splitter is an asynchronous gate that accepts an SFQ pulse and produces two output pulses after its intrinsic delay. One splitter can produce only two fanouts. For additional fanouts, more splitters should be added in a binary tree structure. To have $n$ fanouts, $n$-$1$ splitters are needed. Fig. \ref{Splitter} shows the circuit-level schematic of a splitter gate, its operating waveforms, and two examples of splitter binary trees for providing four fanouts (FO4). Please note that for AQFP, splitters are clocked buffers that can have 1-to-2, 1-to-3 and even 1-to-4 fanouts \cite{xu2017synthesis, narama2016study}. 
\subsubsection{Gate-level pipeline}\label{pipeSubSubSec}Unlike CMOS gates, in SFQ logic, most of the gates receive a clock signal. There are three main methods for clock distribution in SFQ circuits: (i) \textit{counter-flow clocking} where the clock flows in the opposite direction of the data, (ii) \textit{concurrent-flow clocking} in which the clock and data flow in the same direction, and (iii) \textit{clock-follow-data} in which  the clock arrives at a gate after its inputs have arrived and processed by the gate. For more information on clock distribution, please see \cite{friedman2001clock,tadrosrobust_isec}.

\subsubsection{Path Balancing}\label{PathBalancingSubSubSec}For an SFQ gate to operate correctly, all of its fanin gates should have the same logic level. If there is a difference among logic levels of fanins of a gate, some DFFs should be inserted into outputs of fanin gates with smaller logic levels \cite{pasandi2018graph}. For example, if the first fanin ($in_1$) of an $AND2$ gate has a logic level of three and the second fanin ($in_2$) has a logic level of four, one DFF should be added to the output of $in_1$. Without path balancing, correct pulses on $in_1$ will be consumed by this $AND2$ gate one clock before arrival of the corresponding pulses on the second input, hence, this gate will not be able to produce correct output values. 

Input pulses to an SFQ gate can be modeled by ``tokens" that must arrive at the same clock period and should be consumed by the  clock pulse arrives at the end of this period. Path balancing guarantees correct arrival and consumption of these tokens. 
\begin{figure}[t]
\centering
\includegraphics[width=0.4\textwidth]{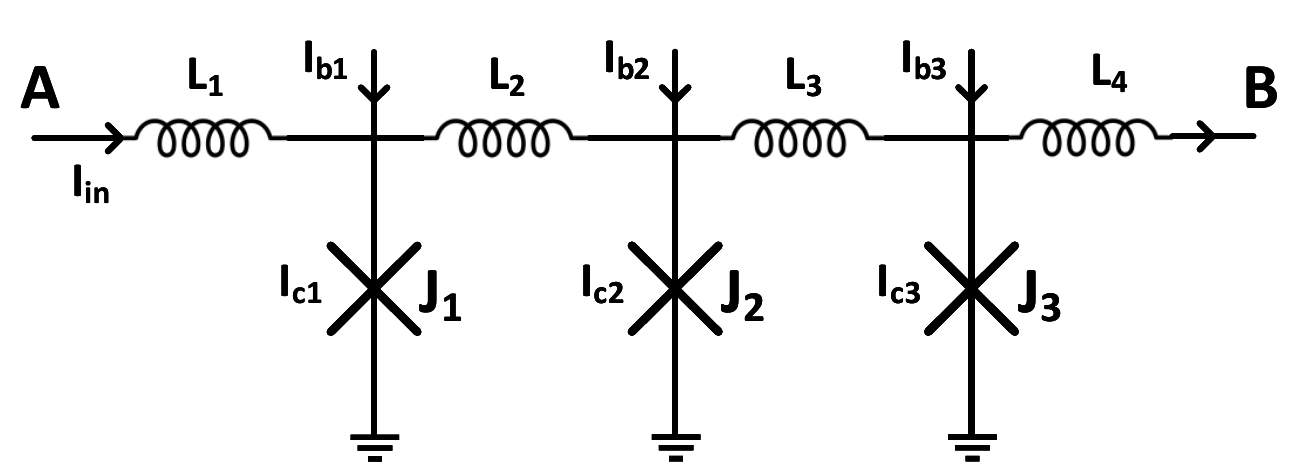}
\caption{Josephson Transmission Line (JTL).}
\label{JTL_fig}
\end{figure}
\subsubsection{Interconnects in SFQ}\label{InterSubSubSec}In SFQ circuits, there are two methods for transmision of signals among gates: using \textit{Josephson Transmission Line (JTL)}, and using \textit{Passive Transmission Line (PTL)}. Most of SFQ circuits use the JTLs for transmitting signals among the gates, mainly because they can transfer the SFQ pulses without any distortion \cite{polonsky1993transmission, hashimoto2003design}. Fig. \ref{JTL_fig} shows the circuit-level schematic of a JTL. This circuit can also provide current and power gains (amplifications). For this purpose, the critical current of JJs should grow in the propagation direction ($I_{c1} < I_{c2} < I_{c3} < ...$), and the inductance values should decrease ($L_1 > L_2 > L_3 > ...$) in that direction \cite{likharev1991rsfq}. The PTLs are similar to microstrip and strip lines. Each PTL requires one transmitter and one receiver gate. Generally, for short interconnects JTLs, and for long interconnects PTLs are used \cite{hashimoto2003design}.
\subsection{Background on Logic Synthesis}
\label{BG-Synth:sec}
Logic synthesis is divided into two phases: \textit{technology-independent} and \textit{technology-dependent} (\textit{technology mapping}) phase. In the first phase, several optimizations are performed to reduce total number of literals in the given network. Some useful operations include: common sub-expression extraction, decomposition, and re-substitution. In the second phase, suitable gates from a given library are assigned to nodes of the given network in order to meet some constraints and/or to minimize some cost functions. Before technology mapping, the given network is transformed into a network of (N)ANDs and inverters. This step is called \textit{technology decomposition}, and the resulting graph is called \textit{subject graph}.

A $k$-$feasible$ $cone$ at node $v$ of a network $N=(V,E)$, denoted by $C_v$, is defined as a sub-graph containing $v$ and its predecessors satisfying two conditions: (i) number of inputs of this sub-graph should be fewer than or equal to $k$, (ii) all paths connecting $v$ to a node in $C_v$ lies entirely in $C_v$. A cut $C$=$(X,X')$ with source $s$ and sink $t$ in a given network $N$ is defined as a partition of $V$ into $X$ and $X'=V-X$ such that $s \in X$, and $t \in X'$. $C$=$(X,X')$ is a trivial cut, if set $X$ has only one member (source $s$). The node cut-size of a cut $C$=$(X, X')$ denoted by $n(X, X')$ or $n(C)$ is defined as the number of boundary nodes in $X'$ which are adjacent to some nodes in $X$. These boundary nodes are called the leaf nodes of the cut.

A cut $C$=$(X,X')$ is called $k$-$feasible$ if its node cut-size is at most $k$ (i.e. $n(C) \leq k$). A $k$-$feasible$ $cut$ of a node $v$ is defined as a valid $k$-$feasible$ cut, in which node $v$ is the source node of the cut and the sink node is a PI. A fanin (fanout) cone of a node $v$ in a network $N=(V,E)$ is defined as the set of nodes in $V$ that can be reached through the fanin (fanout) edges of $v$. Maximum Fanout Free Cone (MFFC) of a node $v$ is a subset of its fanin cone in which any path from a node in this subset to any PO of the network goes through $v$. During optimizations, if a node is removed, all nodes in its MFFC can be removed as well. 

A \textit{binary tree} is a tree in which nodes have either one child or two children. A full or saturated binary tree with height (or depth) $H$ is a binary tree which contains $2^H$-$1$ nodes. A binary tree with height $H$ but with fewer number of nodes is called a general binary tree. The binary tree that we will consider in the rest of this paper is a binary tree in which all nodes have two children; a child can be another node in the tree or a PI.
\subsection{Prior Work}
\label{Prior-Work:sec}
In the literature of the logic synthesis and verification, there are many papers addressing the technology-independent and technology mapping phases. Some of these papers developed useful algorithms/tools and invented effective heuristics for optimizing some objective functions such as literal count \cite{ashenhurst1957decomposition, lawler1964approach, brayton1984logic, brayton1987mis,keutzer1987dagon,brayton1990multilevel} or for increasing the verification speed by presenting fast SAT solvers \cite{moskewicz2001chaff,een2003extensible}. Examples are SIS \cite{sentovich1992sis}, MVSIS \cite{mvsismvsis}, and Chaff \cite{moskewicz2001chaff}. Furthermore, there are many innovative methods such as integration of technology mapping and retiming \cite{pan1998new,mishchenko2006integrating}, or logic decomposition during technology mapping \cite{lehman1997logic}. A logic synthesis and verification tool, ABC \cite{synthesis2011abc}, has been developed by the Berkeley verification and synthesis research group to provide a flexible programming environment to implement the recent innovations. 

In the literature, there are also some papers which present logic synthesis algorithms and tools for some specific applications.  For example, in \cite{tsui1993technology}, authors proposed tree mapping and decomposition algorithms to generate a power efficient mapping solution for a given network. In \cite{vaishnav1995delay,tiwari1993technology,yeh1999technology,tiwari1996technology},  some other technology mapping methods targeting the reduction of power consumption are presented. In \cite{mishchenko2007combinational}, a priority-cut-based technology mapping is presented in which the priority of selecting matches can be set as delay, area, or any other metric. In \cite{chaudhary1992near}, a near optimal algorithm for technology mapping is proposed. This algorithm minimizes area under delay constraints by generating area-delay curves. 

There are a few papers addressing the logic synthesis for SFQ circuits \cite{yamashita2006transduction,yoshikawa2001top,katam2017desig_isec,pasandi2018sfqmap}. In \cite{yamashita2006transduction}, a framework is developed by constructing a virtual cell called ``2-AND/XOR". This framework allows usage of the CMOS logic synthesis tools for SFQ circuits as claimed in \cite{yamashita2006transduction}. In \cite{yoshikawa2001top}, a Binary Decision Diagram (BDD)-based top-down design methodology is used for SFQ circuits. In \cite{katam2017desig_isec}, the required path balancing DFFs and the splitter cells are added to the netlist generated by ABC \cite{synthesis2011abc} followed by applying the standard retiming algorithm \cite{leiserson1991retiming} to reduce the required number of path balancing DFFs. In \cite{pasandi2018sfqmap}, a technology mapping tool for SFQ logic circuits (called $SFQmap$) is presented which provides two main optimizations: (i) logical depth minimization with path balancing, and (ii) peephole optimization for minimizing product of the worst-case stage delay and the logical depth (PSD).

In this paper, we present a path balancing technology mapping algorithm which favors generating mapping solutions with balanced structures. In addition, several closed form formulas are developed which relate the number of leaf nodes in a tree with the required path balancing DFF count at each level of this tree. Thanks to these formulas, the optimality of path balancing tree mapping algorithm is proven for SFQ logic circuits. Path balancing can be considered during different phases including \textit{technology independent optimizations}, \textit{technology decomposition}, and \textit{technology mapping}. In this paper, we focus on the \textit{technology mapping} phase. 
\subsection{State-of-the-art Technology Mapping Flow}
\label{state_mapping_flow}
In the technology mapping flow of the state-of-the-art mappers, as explained in \cite{mishchenko2005technology}, at first the $k$-$feasible$ $cuts$ and cuts' fucntions based on their inputs are computed for each node of the given network. Next, in a topological ordering traversal, and by using Boolean matching \cite{mailhot1990technology}, the best matches and their best implementations using the pre-generated supergates \cite{mishchenko2005technology} are extracted. At the end, the best cover for the given network is generated in a reverse topological ordering traversal. This approach is followed by ABC \cite{synthesis2011abc} as well. In the following, each of these steps are explained in more details.
\subsubsection{Computing k-feasible cuts}
\label{k-feas}
$k$-$feasible$ $cuts$ for nodes in the given network are computed in a way similar to \cite{cong1994flowmap}. For each node, a trivial cut consisting of the node itself is added to the set of cuts. Having this trivial cut and existence of (N)AND and inverter in the library, the feasibility of finding a mapping solution for any given network is guaranteed. 
\subsubsection{Computing Cut's Function}
\label{comp_cut_func}
For all computed $k$-$feasible$ $cuts$, except trivial cuts, the cut's function (sometimes called truth-table) is computed. Function of a trivial cut is the same as the Boolean expression of the source of this cut. The function of a non-trivial cut is computed by assigning some variables to the inputs of the cut. Using these variables, the truth-table of the cut is computed by performing some simulations. One round of simulation includes propagation of a set of inputs through the network \cite{zhang2005simulation}. Some bit-parallel methods such as what is presented in \cite{baeza1992text} is used to increase the speed of simulations. Next, the function of a cut is stored inside a field in the data structure of this cut. Since usually $4-6$-$feasible$ $cuts$ are considered, a variable with length of $16-64$ bits is enough for storing the function of a cut.
\subsubsection{Supergates}
\label{superg:sec}
A supergate is a small single-output combinational network composed of the original gates in the given library. Supergates are generated by exhaustively concatenating the original gates in the library. This is done as a pre-processing step after reading the library and before perfoming the technology mapping. Generation of a supergate is controlled by the following factors: number of inputs of the supergate, total run-time for generating all supergates, area of the supergate, critical path delay of the supergate, and the depth of the supergate. Other than addressing the structural bias problem \cite{mishchenko2005technology} by looking deeper into the network, using supergates makes the cut-enumeration-based method for library-based technology mapping more reasonable by providing implementation choices for more cuts. 
\subsubsection{Boolean Matching}
\label{Boolean_match}
In the state-of-the-art technology mapping flow, Boolean matching \cite{mailhot1990technology,benini1997survey} is used to identify whether the Boolean function of a cut can be implemented using generated supegates. A hash-table of functions of supergates is produced and during the mapping phase, function of a cut is looked up in this hash-table in a constant time.
\subsubsection{Best Matches and Best Covers}
\label{best_m_c}
\begin{table}[t]
  \centering
  \caption{Average hit rate for 20 ISCAS benchmark circuits in the standard cut-enumeration-based technology mapping approach \cite{synthesis2011abc}.}
    \begin{tabular}{cccc}
    \toprule
      & \multicolumn{3}{c}{Supergate Level}\\
    \toprule
       k   & $L = 1$ & $L = 2$ & $L = 3$ \\
    \midrule
	5 &0.056 & 0.109 & 0.118 \\
	\midrule
	6 & 0.047 & 0.057 & 0.067 \\
	\bottomrule
    \end{tabular}%
  \label{hitrate_table}%
\end{table}%
In a topological ordering traversal, the best matches for both phases (positive and negative) of each node for its best cut is computed and a supergate that gives the least arrival time for that cut is selected. After finishing this traversal and reaching the POs, in another traversal in a reverse order, best supergates for implementing functions of gates connected to POs are selected. Next, best gates implementing inputs of those supergates are chosen and so on. After the PIs of the network are visited, a mapping solution for the entire network is generated.

Cut-enumeration-based technology mapping with using $k$-$feasible$ cuts suits best for $LUT$-based FPGA technology mapping as in \cite{cong1994flowmap,cong2003optimal}. This is because for any computed $k$-$feasible$ cut, there will be a $k$-$LUT$ that can implement the function of this cut. However, for library-based technology mapping, theoretically, most of the time there will not be any gates in the library to implement the function of a cut. For example, by having 20 gates in the original library, and using up to level 3 supergates, there will be around 4000 supergates in the supergate library. Thus, the probability of having a supergate to implement the function of a cut in the set of $k$-$feasible$ cuts for $k=5$ and $k=6$ will be $\frac{4000}{2^{2^5}} \approx 10^{-6}$, and $\frac{4000}{2^{2^6}} \approx 10^{-15}$, respectively. In these calculations, the fact that there is at most $2^{2^k}$ different $k-$input functions is employed. In this paper, the aforementioned probabilty is called the hit rate\footnote{The ratio of the total number of cuts that have at lease one supergate in the supergate library capable of implementing their function to the number of cuts without having any supergates that can implement their functions.}. Fortunately, for most of the practical circuits, there will be much fewer number of cuts than the upper bound of $2^{2^k}$. This results in having much better values for the hit rate. Table \ref{hitrate_table} shows the average hit rate  of 20 ISCAS \cite{hansen1999unveiling} benchmark circuits for two values of $k$ and three levels of supergates. As seen, the practical hit rate, specially for $L=3$, is much better than the aforementioned theoretical worst case value. Therefore, it is reasonable to use $k$-$feasible$ cuts together with supergates for library-based technology mapping. In our developed technology mapping tool, the state-of-the-art flow is followed.
\section{Proposed Path Balancing Technology Mapping Algorithm}
\label{DP_Tech-Map}
\begin{figure}[t]
\centering
\includegraphics[width=0.5\textwidth]{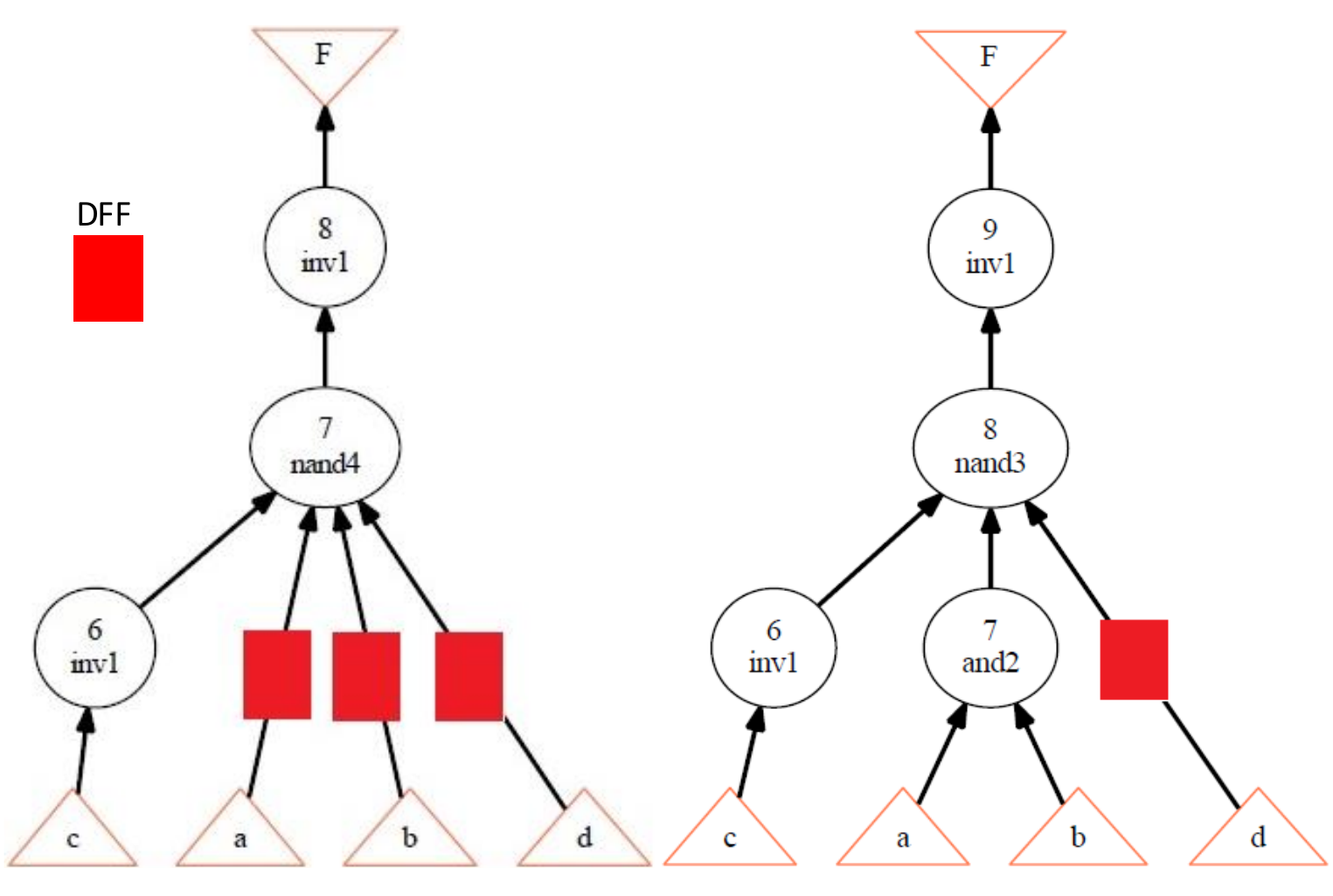}
\caption{Two mapping solutions for $F = a.b.(!c).d$. The left circuit, generated by ABC's mapper \cite{synthesis2011abc} and requires three path balancing DFFs. There is another mapping solution with only one DFF as shown in the right graph.} 
\label{Motive_example}
\end{figure}
\subsection{Motivation}
\label{Motiv:sec}
Suppose that we want to map the following expression: $F = a.b.(!c).d$. As shown in Fig. \ref{Motive_example}, state-of-the-art mappers (such as ABC \cite{synthesis2011abc}) produce the left circuit which requires three path balancing DFFs. However, it is possible to have a better mapping solution with fewer number of required path balancing DFFs, as shown in the right graph in Fig. \ref{Motive_example}. This is because there is no implemented algorithm in the current state-of-the-art technology mappers for controlling balancing of the network which is being mapped. In the next section, we present a novel path balancing technology mapping approach which generates mapping solutions with minimum number of required path balancing DFFs for mapping trees. From now on, we denote the total number of required path balancing DFFs by \textit{\#DFFs}. 
\subsection{Presenting Our Algorithm}
\label{presenting-DP:subsec}
We present the problem of path balancing tree mapping as a dynamic programming (DP) problem. The input of the technology mapper is a network of two input (N)AND and inverters which is called the subject graph. ABC uses And Inverter Graphs (AIGs) to represent subject graphs. In AIGs, all nodes are two input AND gates. Inverter is modeled as a field in the data structure of the node. Therefore, if the subject graph is a tree, it can be modeled as a binary tree in which all nodes have two children. A child can be a node or a PI. The goal is to find a mapping solution for the given subject graph with fewest \#DFFs.

In the path balancing technology mapping algorithm, the optimal solution for mapping a tree rooted at node $v_i$ is defined as a solution which minimizes \#DFFs. Suppose that the set of all $k$-$feasible$ cuts of node $v_i$ is $\Bbb K_i$, and for a $k$-$feasible$ cut $C_j \in \Bbb K_i$, $L_{C_j}$ denotes its set of leaf nodes (inputs).  The value of the optimal solution, $OPT(v_i)$, is calculated recursively using the following equation:
\begin{equation}
OPT(v_i) = min \left \{ {\sum_{\forall v \in L_{C_j}}OPT(v)} \quad + \ B(L_{C_j}) \right \} \forall C_j \in \Bbb K_i
\label{OPT_i}
\end{equation}
\begin{figure}[b]
\centering
\includegraphics[width=0.4\textwidth]{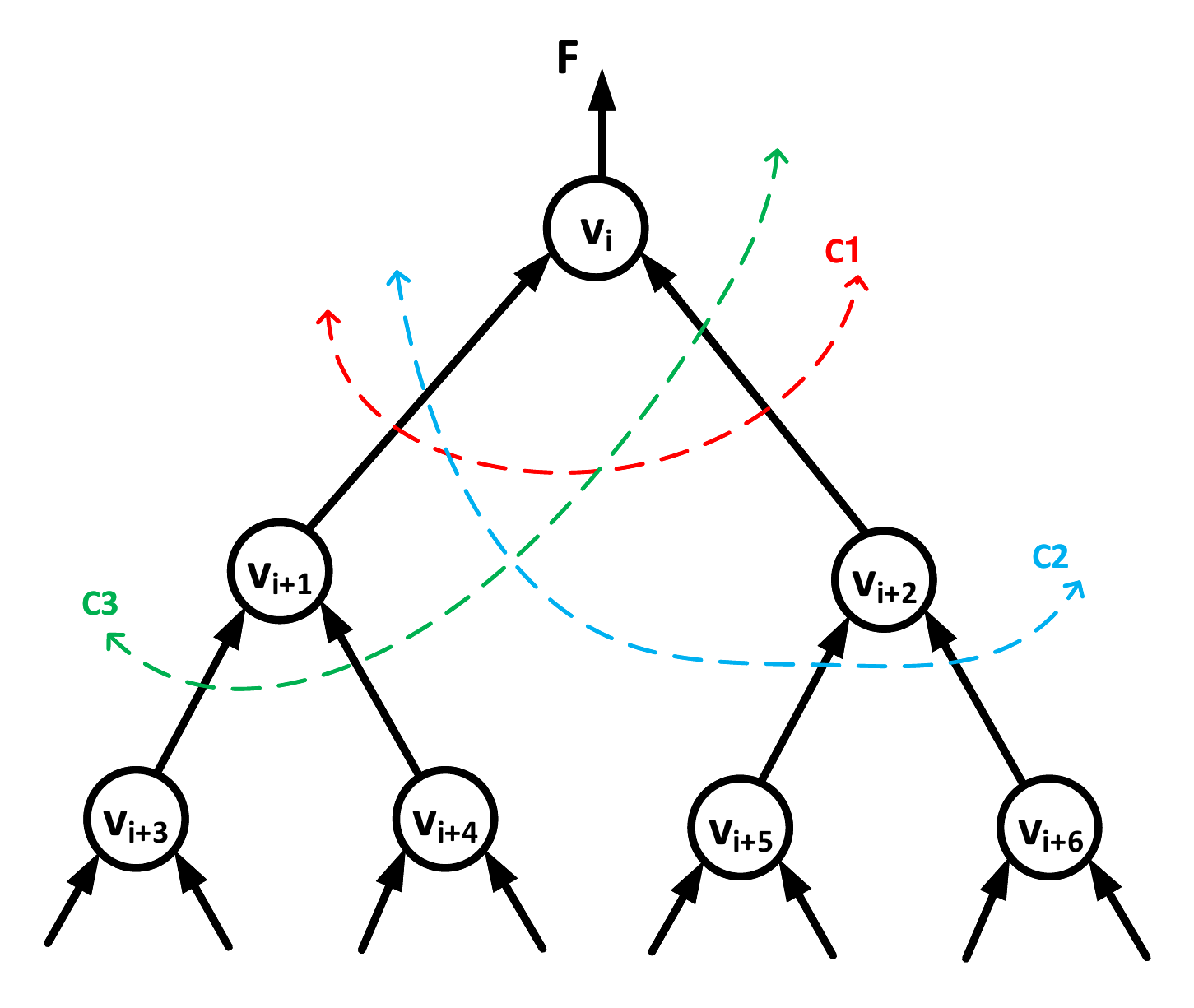}
\caption{Showing 3-feasible cuts of node $v_i$.}
\label{3-feasible-cuts}
\end{figure}
in which, $B(L_{C_j})$ is a function which receives the set of leaf nodes of a cut ($C_j$ here) and returns the required number of DFFs for balancing the inputs of this cut. This balancing is required if there is a difference among logic levels of inputs of the cut. For example, suppose that $C_1$ has two leaf nodes $v_1$ and $v_2$ ($L_{C_1}=\lbrace v_1,v_2 \rbrace$). If levels of $v_1$ and $v_2$ are three and five, respectively, $B(L_{C_1})$ will return two.

\textit{Example}: Consider the binary tree shown in Fig. \ref{3-feasible-cuts}. The $3$-$feasible$ cuts of node $v_i$ are shown in this figure. Using Eq(\ref{OPT_i}) and having $k$=$3$, the value of the optimal solution for node $v_i$ is computed as follows:
{\small
\begin{align}
& OPT(v_i) = min \lbrace \nonumber\\
& OPT(v_{i+1}) + OPT(v_{i+2}) + B(\lbrace v_{i+1},v_{i+2} \rbrace)\textbf{,} \nonumber\\
& OPT(v_{i+1}) + OPT(v_{i+5}) + OPT(v_{i+6}) + B(\lbrace v_{i+1},v_{i+5},v_{i+6} \rbrace)\textbf{,} \nonumber\\ 
& OPT(v_{i+2}) + OPT(v_{i+3}) + OPT(v_{i+4}) + B(\lbrace v_{i+2},v_{i+3},v_{i+4} \rbrace)\textbf{,}\nonumber\\
& \rbrace
\label{OPT_i_example}
\end{align}
}
The optimal path balancing tree mapping solution is generated as follows:\\
In a topological ordering traversal starting from level-1 nodes, the $k$-$feasible$ cuts for each node in a way similar to \cite{cong1994flowmap}, and each cut's function based on its inputs in a way similar to \cite{mishchenko2005technology} are computed. Afterwards, the best valid solution for each node which minimizes \#DFFs is found in a DP approach using Eq(\ref{OPT_i}). The tree traversal is continued until the root of the tree is visited. After visiting the root, the optimal path balancing mapping solution for the whole tree is calculated. This solution can be generated by tracing the tree from its root all the way back to its PIs. 
We will prove that the presented DP based algorithm for path balancing technology mapping provides optimal solutions for mapping trees when an SFQ library of gates is used. 

The complexity of computing the $k$-$feasible$ cuts is $O(kmn)$, where $m$ is the edge count, and $n$ is the node count \cite{cong1994flowmap}. The complexity of computing cut functions is linear in the size of the network \cite{zhang2005simulation}. By having the $k$-$feasible$ cuts, the complexity of the path balancing tree mapping algorithm is $O(K'ng)$, in which $K'$ is the maximum number of $k$-$feasible$ cuts for any node of the subject tree, $n$ is the node count, and $g$ is the number of supergates in the supergate library. The overall complexity of the algorithm is $O(kmn)$.

In order to use DP for finding the optimal solution for a problem, this problem should satisfy the DP's principle of optimality. For this purpose, the optimal solution to the problem should be built of the optimal solutions to its sub-problems. It looks like it is possible to find some examples in which the optimal path balancing mapping solution for a tree rooted at node $v_i$ is not built of the optimal solutions to its sub-problems. For node $v_i$ in Fig. \ref{3-feasible-cuts}, assume that its best cut is $C1$, and suppose that for a tree rooted at node $v_{i+2}$, there is a single match of $(7,4)$, and for node $v_{i+1}$, there are two matches $(3,2)$ and $(5,3)$. A match is shown with a couple $(x,y)$. The first attribute stands for the height or depth of the match, and the second attribute stands for \#DFFs.
The best mapping solution for node $v_i$ will contain $(5,3)$ for node $v_{i+2}$. This is because it gives $3$+$4$+$(7$-$5)$ = $9$ required path balancing DFFs, while the other mapping solution for $v_{i+2}$ gives $2$+$4$+$(7$-$3)$ = $10$ required path balancing DFFs. Therefore, in this scenario, the best mapping solution for node $v_i$ is not built of the best (with the least \#DFFs) for node $v_{i+2}$. This means that this example disproves the DP's principle of optimality for path balancing tree mapping. 

We will prove that these kinds of counter examples do not exist in actual circuits. For this purpose, we need to prove that by increasing the height of a sub-tree from $H$=$X$ to $H$=$X$+$p$, \#DFFs for internal balancing of the sub-tree will be increased by more than p, where p is a natural number.
Unfortunately, by having gates with $k>2$ inputs in the library, the problem becomes very complicated. In the following, we provide a proof of optimality for the case of having gates with only two inputs in the library. This is valid for the used SFQ library of gates \cite{RSFQLib}. In the proof, it is needed to have a closed form formula for the total number of input pins of the mapped tree based on \#DFFs for that tree. This formula is developed in Section \ref{proof:sec}.

We use \#DFFs as a metric for measuring how balanced a graph is; if a graph has smaller value for \#DFFs, this means that it is more balanced. Therefore, minimizing \#DFFs for a graph during technology mapping results in achieving the most balanced mapping solution for this graph. 
\subsection{Terminology}
\label{term:sec}
\begin{figure*}[t]
        \centering
        \begin{subfigure}[!t]{0.42\textwidth}
                \centering
                \includegraphics[width=\textwidth]{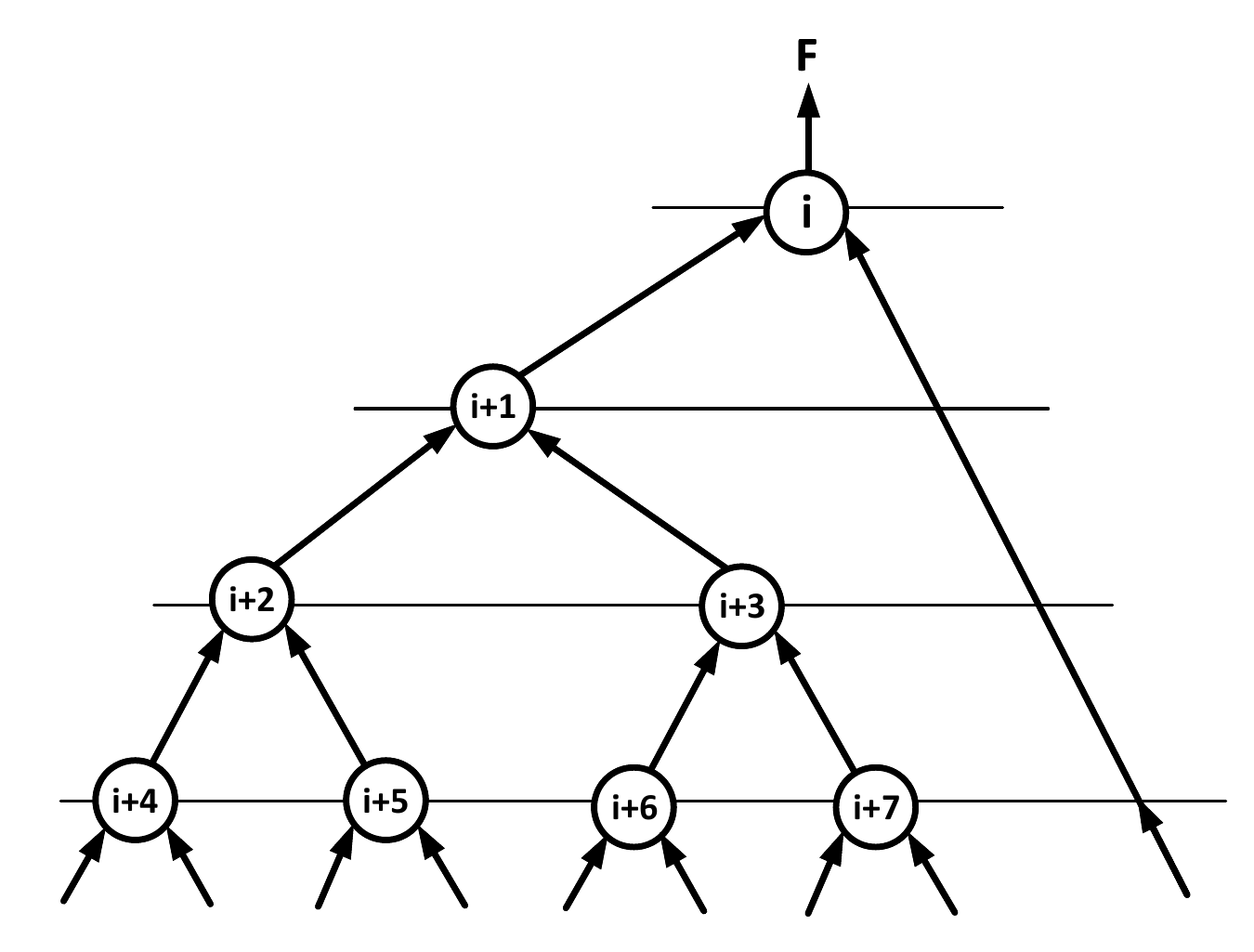}
                \caption{}
                \label{ret1}
        \end{subfigure}
        \begin{subfigure}[!t]{0.5\textwidth}
                \centering
                \includegraphics[width=\textwidth]{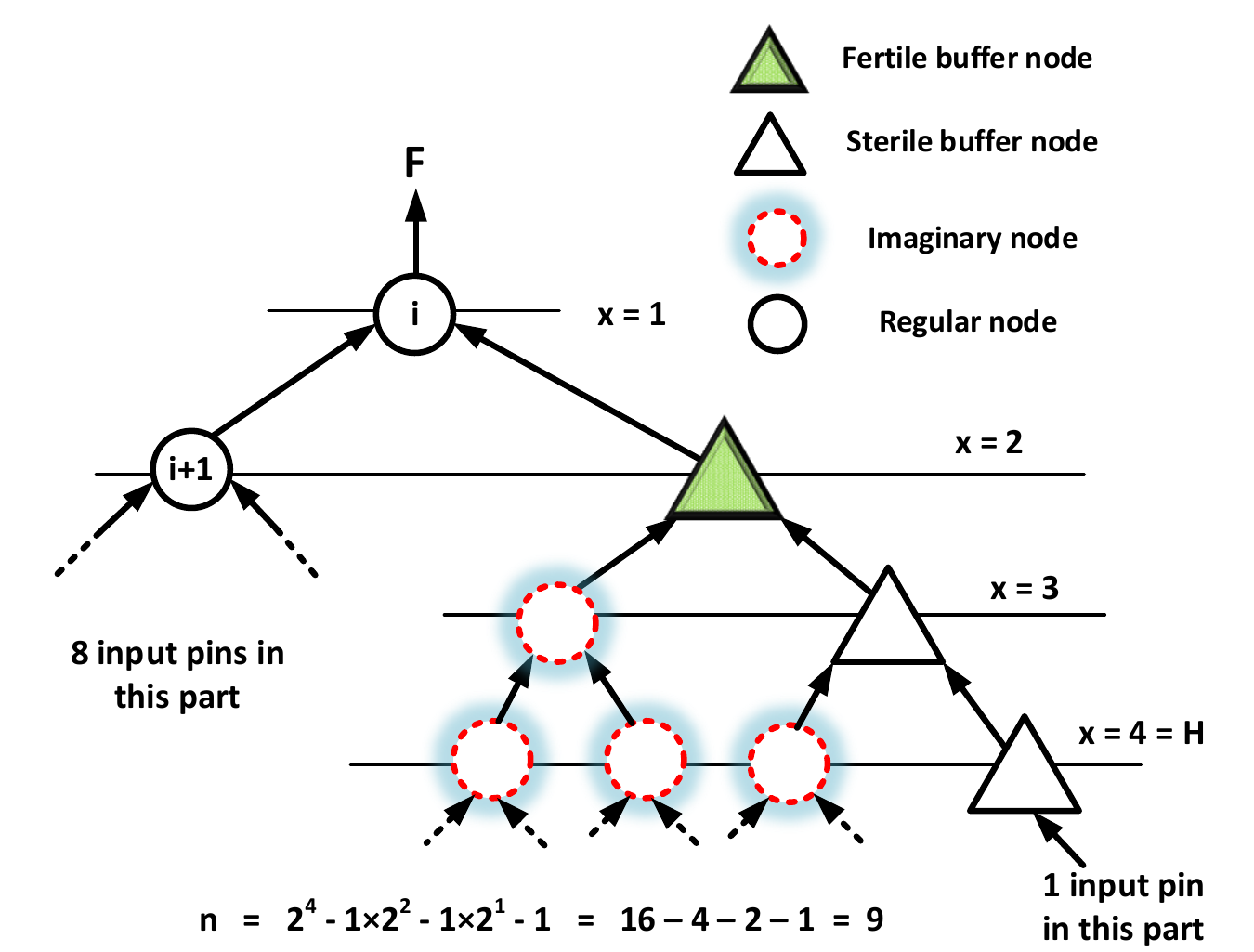}
                \caption{}
                \label{ret2}
        \end{subfigure}
        \caption{(a) A tree that we want to find its \#input pins, and (b) its extended tree. Fertile and sterile buffer nodes and imaginary nodes are shown in the extended tree. The left sub-tree (not shown) in the extended tree is a full binary tree rooted at node $i$+$1$. This sub-tree generates $2\times4$ input pins, and there is a single input pin feeding the only buffer node at the last level ($x$=$4$). Thus, $n$=$9$.}
        \label{gen_example}
\end{figure*}
\textbf{input pin}: Primary input (or leaf node) of a tree. \\
\textbf{inputs} vs \textbf{input pins} : The first one is used to refer to fanins of a gate, while the second one is used to refer to leaf nodes of a tree. \\
\textbf{$n$}: Total number of input pins of a tree.\\
\textbf{\#input pins}: Total number of leaf nodes (or PIs) of a tree ($=n$). \\
\textbf{$N$}: Total number of internal nodes of a tree. \\
\textbf{x}: Level of a node in a tree. Root of the tree is at level one and levels of other nodes are higher than one.\\
\textbf{H}: Height of a tree (last level of a tree, furthest from the root). \\
\textbf{Buffer Node}: A node which has one father node and one child node. A DFF sits on the place of a buffer node.\\
\textbf{Imaginary Node}: A node which does not have a father node, and actually does not exist in the tree (it is only a concept). This concept is used for calculating \#input pins of a tree.\\
\textbf{Sterile Buffer Node}: A buffer node which is not capable of generating imaginary nodes.\\
\textbf{Fertile Buffer Node}: A buffer node which is capable of generating imaginary nodes. A fertile buffer node also generates one sterile buffer node per level starting from its higher level to the last level of the tree.\\
\textbf{Extended Tree}: A tree obtained by adding all buffer nodes and imaginary nodes to the original tree. Extended tree is only a conceptual thing and it is used for model development. We do not actually construct an extended tree during the technology mapping.\\
\textbf{Generation of imaginary nodes}: Each fertile buffer node generates some imaginary nodes at the higher numbered levels (further down from the root). We are interested in the number of imaginary nodes generated at the last level of the tree. An imaginary node belongs to a buffer node with smallest level (closer to the root) that can be reached from this imaginary node in the extended tree. \\ 
\textbf{$y_i$}: Total number of buffer nodes at level $i$. Sum of all $y_i$s in a mapped tree is equal to the total number of required path balancing DFFs for that tree.\\
\textbf{Y}: Total number of required path balancing DFFs in a mapped tree. $Y = \sum_{i=2}^{H}y_i$. 
\subsection{Discussion about the Algorithm}
\label{proof:sec}
Total number of nodes at the last level ($x$=$H$) of a full binary tree is equal to $2^{H-1}$, in which $H$ is the height of the tree. Since all nodes have two inputs (Section \ref{presenting-DP:subsec}), thus, \#input pins for a full binary tree is $2^{H}$. A general binary tree has fewer number of internal nodes, fewer nodes in the last level, and fewer \#input pins compared with a full binary tree. In the following, a closed form formula for \#input pins of a general binary tree is developed.

In a general binary tree, there are some missing nodes compared with a full binary tree with the same height; wherever there is a missing node, a buffer node will sit in that place. If this node (at level $x$) was not missing, it could create $2\times2^{H-x}$ input pins at the last level of the tree. So, this amount of \#input pins should be deducted from \#input pins of a full binary tree to achieve \#input pins for this general binary tree. This contributes to the reduction of the \#input pins by $2\times2^{H-x}$-$1$. The `-1' is due to the fact that each chain of buffer nodes, starting from a fertile buffer node all the way down to the last level, needs one input pin. We should be careful about not over-counting the number of fertile buffer nodes. Referring to the definition of the fertile and sterile buffer nodes, if the total number of buffer nodes at level $x$+$1$ is the same as level $x$, it means no new fertile buffer node is generated at level $x$+$1$. Therefore, the total number of fertile buffer nodes at level $x$+$1$ is $y_{x+1}-y_{x}$.

Based on the above discussion, we can write the following formula for \#input pins ($n$) of a general binary tree with height $H$:
\begin{multline}
n = 2^H - y_2 \times 2^{H-1} - (y_3-y_2) \times 2^{H-2} - (y_4-y_5) \times 2^{H-3} - \\
... - (y_{H-1}-y_{H-2}) \times 2^{2}-(y_H-y_{H-1}) \times 2^1 + y_H
\label{n_formula}
\end{multline}

By performing some simplifications on Eq(\ref{n_formula}), the final closed form formula for \#input pins of a binary tree will be as follows:
\begin{multline}
n = 2^H - y_2 \times 2^{H-2} - y_3 \times 2^{H-3} - y_4 \times 2^{H-4} \\
- ... - y_{H-1} \times 2^{1} - y_H
\label{n_formula2}
\end{multline}

Fig. \ref{gen_example} shows a tree, its extended tree, and displays the calculation of \#input pins for this tree using Eq(\ref{n_formula2}) and the concepts of buffer nodes and imaginary nodes. Note that eventhough an imaginary node is connected to the buffer node at $x$=$3$, it belongs to the buffer node at $x$=$2$ based on the definitions and terminology presented in Section \ref{term:sec}. Therefore, the buffer node at level $x$=$3$ is not considered fertile.

For future use, we rearrange the above equation to obtain the following one:
\begin{multline}
y_H + y_{H-1} \times 2^{1} + ... +  y_4 \times 2^{H-4} +  y_3 \times 2^{H-3} +  y_2 \times 2^{H-2} \\
= 2^H - n
\label{n_formula3}
\end{multline}

Now we are ready to present required lemmas and the main theorem in order to prove optimality of the algorithm presented in Section \ref{presenting-DP:subsec}. From now on, we use tree and binary tree interchangeably. 

\textbf{Lemma 1}: Total number of input pins ($n$) for a binary tree is one more than the total number of nodes in this tree, i.e., $n$=$N$+$1$. Recall that in our problem, all nodes of a binary tree have two children.\\
\textbf{Proof}: Please see Appendix \ref{appendix_a}. 

Suppose that there is a binary tree $t_1$ with height $X$ and \#input pins of $n$. Suppose that the height of this tree is increased from $X$ to $X$+$p$ while \#input pins remains the same. The resulting tree is called $t_2$.

\textbf{Theorem 1}: If the total number of buffer nodes of the binary tree $t_2$ is more than the total number of buffer nodes of the binary tree $t_1$ by a positive integer value $\Delta y$, then $\Delta y \geq p$ ($p$ is a natural number). 

Generally, going from a binary tree with height $X$ to a binary tree with height $X$+$p$, while preserving \#input pins, the total number of buffer nodes will be increased. This is because total number of nodes will be the same for both trees (lemma 1), thus, we need to remove at least $p$ nodes from the internal nodes of the first tree and put one at each level to increase the height of the tree from $X$ to $X$+$p$. There are different trees with height $X$+$p$ and \#input pins of $n$. We want to prove that if the total number of buffer nodes (for the new tree with height $X$+$p$) is increased, it cannot increase by less than $p$. So, a valid assumption is considering the most balanced tree for $t_2$, and the least balanced tree for $t_1$. If the theorem is proven for this case, then obviously it is valid for all other cases. 

First, we need to find a lower bound for the total number of buffer nodes of a tree with height $X$+$p$ and \#input pins of $n$, and also an upper bound for the total number of buffer nodes of a tree with height $X$ and \#input pins of $n$. 

\textbf{Lemma 2}: The maximum value for $p$ (difference between height of $t_2$ and $t_1$) in the \textit{Theorem 1} is $n$-$1$-$X$, ($p\leq n$-$1$-$X$). The minimum value is 1. \\
\textbf{Proof}: Please see Appendix \ref{appendix_b}. 

Please note that similar to what is mentioned before, in the following, the most and the least balanced binary trees are ones with the minimum and maximum values for total number of buffer nodes, respectively.

\textbf{Lemma 3}: The most unbalanced (least balanced) binary tree with height $1 \leq X \leq 3$ has $X$ nodes and total number of buffer nodes equals $Y$ = $(X$-$1)X/2$.\\
\textbf{Proof}: Please see Appendix \ref{appendix_c}. 

\textbf{Lemma 4.} The most unbalanced (least balanced) binary tree with height $X\geq 4$ has $2X$-$1$ nodes and $Y$=$(X$-$2)(X$-$1)$ total buffer nodes.\\
\textbf{Proof}: Please see Appendix \ref{appendix_d}.

\textbf{Lemma 5.} The most balanced binary tree with height $X$ and \#input pins of $n$ is obtained as follows:\\
Starting with $y_2$ and maximizing it (=1). If the left hand side of Eq(\ref{n_formula3}) is not larger than its right hand side, we keep $y_2$=$1$ and go for maximizing $y_3$. Otherwise, $y_2$=$0$. Continuing this way and choosing the maximum valid values for $y_i$s that satisfy Eq(\ref{n_formula3}), the resulting tree will be the most balanced tree with height $X$ and \#input pins of $n$.\\
\textbf{Proof}: Please see Appendix \ref{appendix_e}. 

\textbf{Lemma 6.} The most balanced binary tree with height $X$+$p$ and \#input pins of $n$ that can be generated from the most unbalanced binary tree with height $X$ and the same \#input pins (the tree in lemma 4) has $(X$-$p$-$1)(X$-$p$-$2)/2$+$2pX$+$p$-$2p^2$ total buffer nodes.\\
\textbf{Proof}: Please see Appendix \ref{appendix_f}.

Now, we are ready to prove the Theorem 1. By transforming the statements in Theorem 1 into the mathematical expressions, we basically need to prove that the following inequality is not valid for any natural number for $p$: $1 < Y_{diff} <p$, in which, $Y_{diff}$ is the difference between the total number of buffer nodes of the second tree ($t_2$) and the first one ($t_1$). The following equation shows the expression for $Y_{diff}$:
\begin{multline}
Y_{diff} = \lbrace(X-p-1)(X-p-2)/2 + 2pX-p-2p^2 \rbrace -\\
 \lbrace (X-2)(X-1)\rbrace \\
 = (-X^2+4(p+1)X-2p-3p^2-3)/2 
\label{Y_diff}
\end{multline}
in which, $X,p \geq 1$ are natural numbers. Another constraint, as mentioned in the previous lemmas, is $1 \leq p \leq X-1$. By solving these inequalities, it is easy to see that there is no valid values for $p$ that satisfies all inequalities.$\blacksquare$

What we just proved has the following meaning: having gates in the library with no more than two inputs (as in the SFQ library of gates \cite{RSFQLib}), it is not possible to provide a counter example to disprove the optimality of the dynamic programming based approach presented for path balancing tree mapping. Therefore, the path balancing tree mapping algorithm presented in Section \ref{presenting-DP:subsec} gives the optimal solution for 2-input gates and serves as an effective heuristic for multi-input gates.
\begin{figure}[t]
        \centering
        \begin{subfigure}[!t]{0.35\textwidth}
                \centering
                \includegraphics[width=\textwidth]{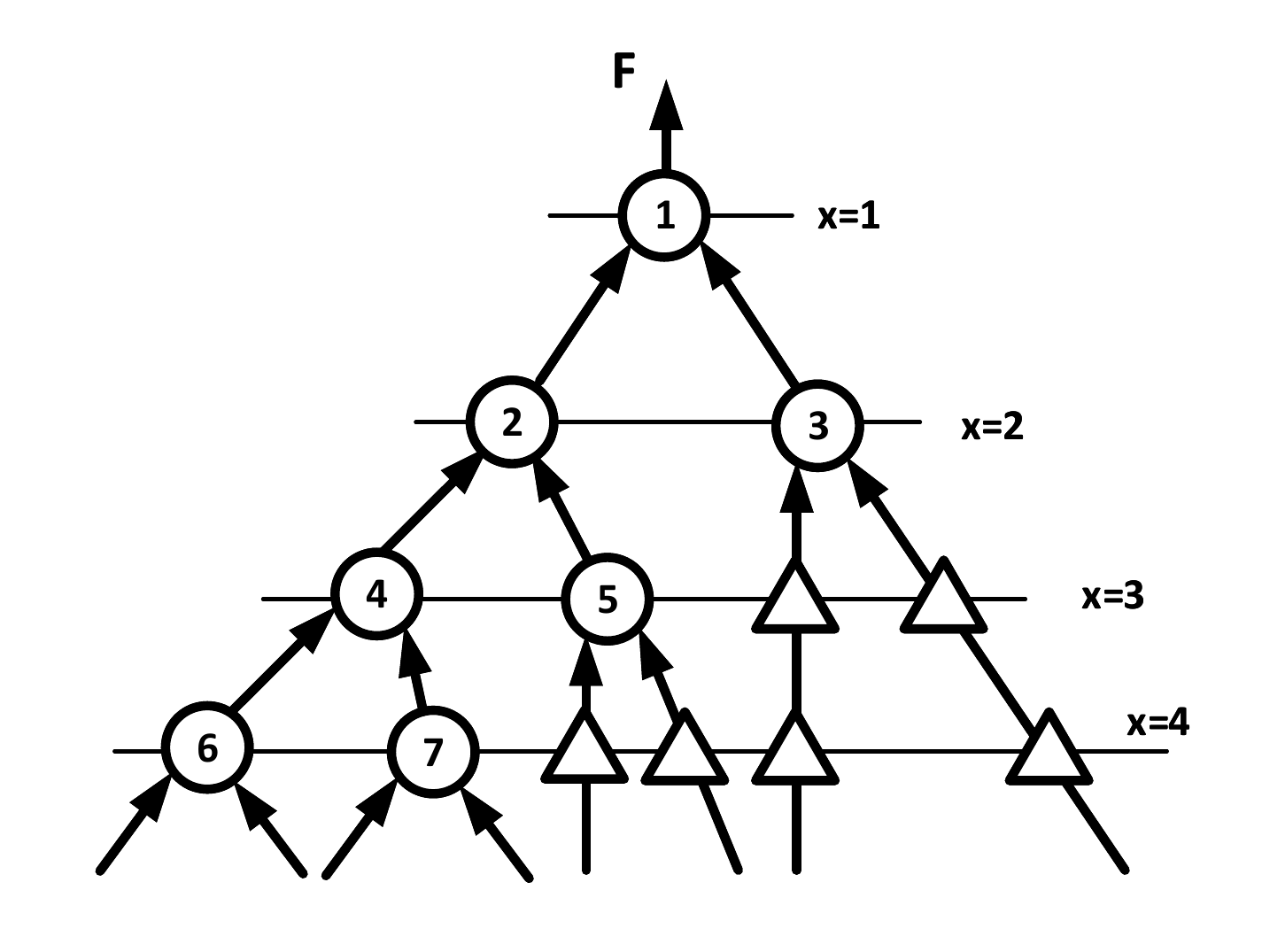}
                \caption{}
                \label{ret1}
        \end{subfigure}
        \begin{subfigure}[!t]{0.35\textwidth}
                \centering
                \includegraphics[width=\textwidth]{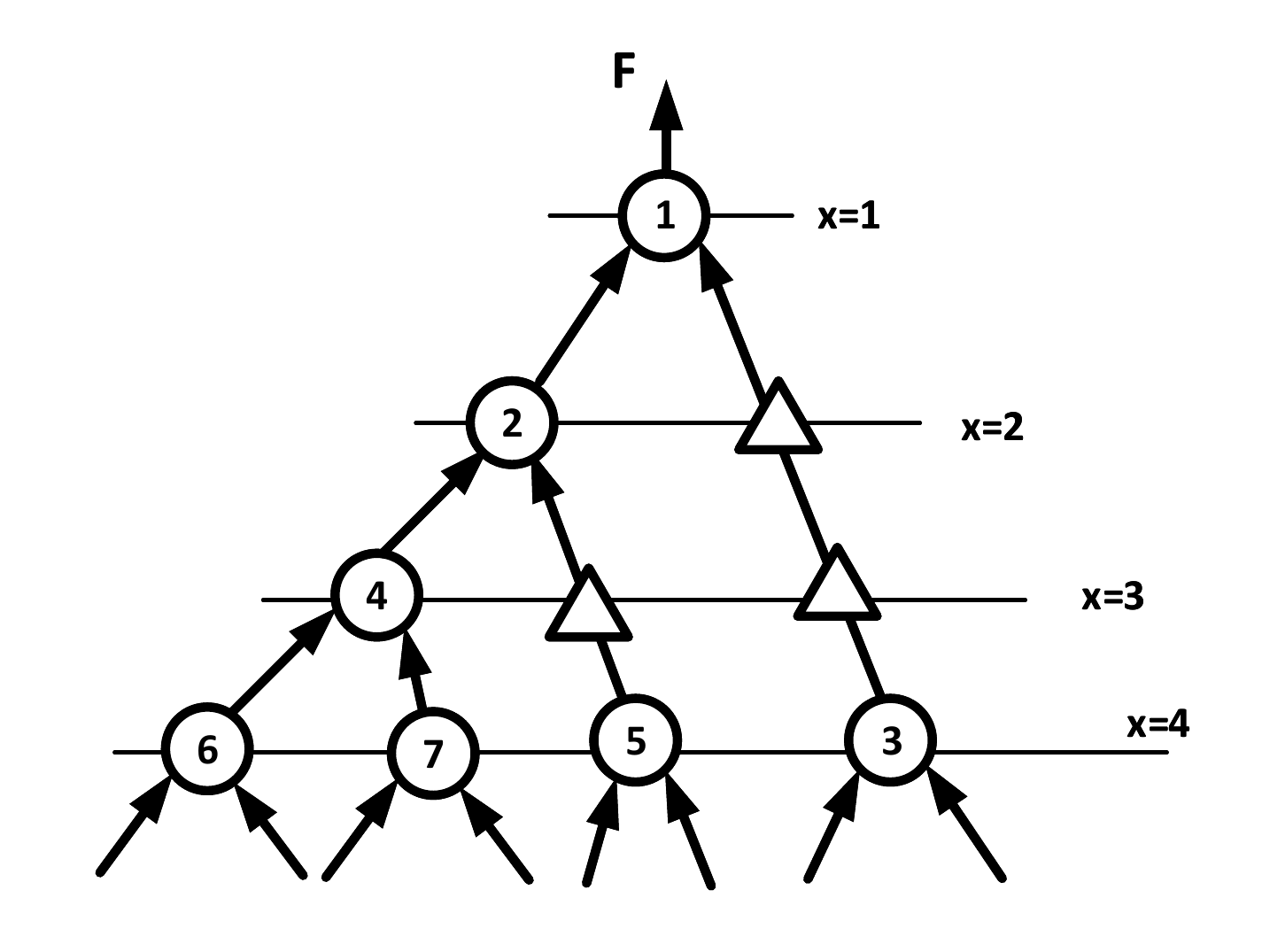}
                \caption{}
                \label{ret2}
        \end{subfigure}
        \caption{Matches for a node in a subject graph: (a) before retiming (b) after retiming.}
        \label{retiming_examples}
\end{figure}
\subsection{Retiming}
\label{retiming:sec}
After finishing the technology mapping and inserting the path balancing DFFs, a standard retiming algorithm \cite{leiserson1991retiming} as in \cite{katam2017desig_isec} can be used to reduce the total number of path balancing DFFs. In our path balancing technology mapping algorithm, we considered the retimed versions of matches for each node during the tree traversal. In other words, the number of path balancing DFFs that is considered in Eq(\ref{OPT_i}) is for retimed matches. Fig. \ref{retiming_examples} shows a match for a node in a subject graph before and after applying the retiming algorithm. In our path balancing technology mapping algorithm, the retimed version (Fig. \ref{ret2}) is used for counting the number of DFFs for a match. Therefore, it should be proven that the developed formulas in Section \ref{proof:sec} for \#input pins is valid for retimed matches as well. For this purpose, it is enough to show that after applying the retiming algorithm to a match, Eq(\ref{n_formula2}) will be valid for relating its \#input pins to its buffer node count.

\textbf{Lemma 7:} Eq(\ref{n_formula2}) is valid for a retimed match. \\
\textbf{Proof}: Please see Appendix \ref{appendix_g}. 
\subsection{DAG Mapping}
\label{DAG_Mapping}
For finding path balancing mapping solutions for DAGs, a cut-enumeration-based method similar to what is presented in \cite{cong1994flowmap,chen2004daomap} followed by a dynamic programming approach similar to what is discussed in Section \ref{presenting-DP:subsec} is used. The subject graph in this case is a DAG as opposed to Section \ref{presenting-DP:subsec} which was a tree. As experimental results in Section \ref{exper:sec} shows, this method provides considerable improvements in reduction of the total number of path balancing DFFs and total area compared with the state-of-the-art technology mappers. Note that for most of the benchmark circuits the subject graph is actually a DAG.
\begin{algorithm} [t]
\caption{PBMap}\label{PBMap_alg}
\DontPrintSemicolon 
\KwIn{Given network: $N=(V,E)$ \\
 }
\KwOut{Mapped network with minimum path balancing overhead: $N_{Map}$}
//pre-mapping computations: \\
Computing $k-feasible$ cuts for each node. \\
Computing $truth-tables$ for each cut. \\
Constructing and initializing the mapping manager, $pMan$. \\
Generating the library of $supergates$.  \\
\For{each node $v$ in $N$}{
 Find the best mapping solution based on Eq(\ref{OPT_i}).
}

Depth\_Minimization ($pMan,N$) \\
Area\_Optimization ($pMan,N$) \\
//generating the mapped network: \\
$N_{Map}$ = Network\_From\_Map ($pMan, N$)

\Return{$N_{Map}$}\;

\end{algorithm}
\subsection{Clock Jitter Accumulation}
Clock jitter accumulation is a measurement of the timing uncertainty at the user defined time offset over the course of a few clock cycles \cite{hollis2014invited}. In the worst case, this can result in obtaining erroneous outputs. Therefore, it is crucial to design a clock distribution network with acceptable amount of accumulated jitter. We believe that in our proposed path balancing technology mapping and retiming algorithm, to a first order, the accumulated clock jitter along input-output paths of the circuit will not be changed compared to conventional path balancing methods. This is because our algorithm reduces the number of required path balancing DFFs for a given circuit without changing the gate-level wave-pipelined structure of the circuit. 
\begin{figure*}[t]
        \centering
        \begin{subfigure}[!t]{0.77\textwidth}
                \centering
                \includegraphics[width=\textwidth]{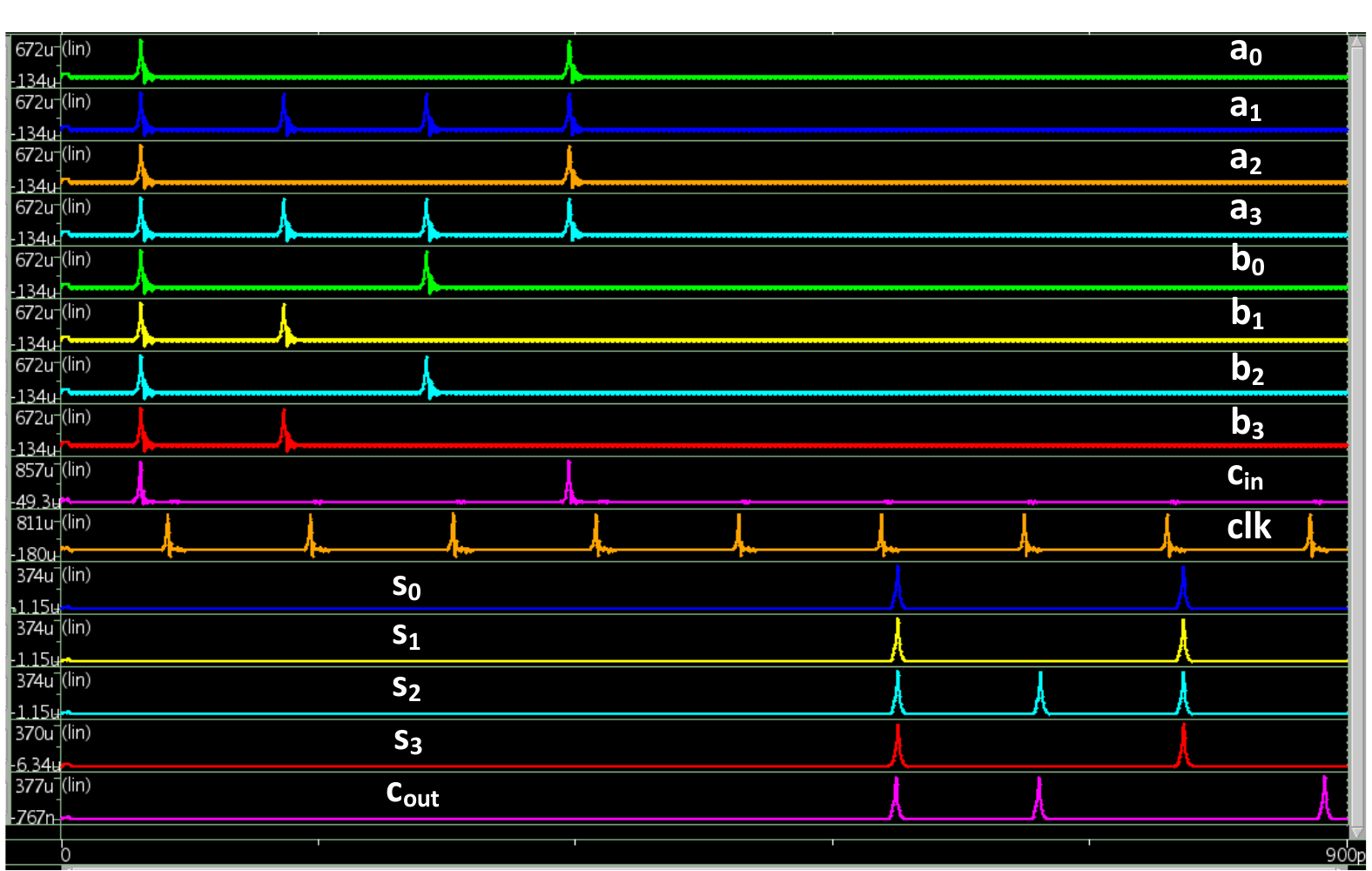}
                \caption{}
                \label{sim_1}
        \end{subfigure}
        \begin{subfigure}[!t]{0.77\textwidth}
                \centering
                \includegraphics[width=\textwidth]{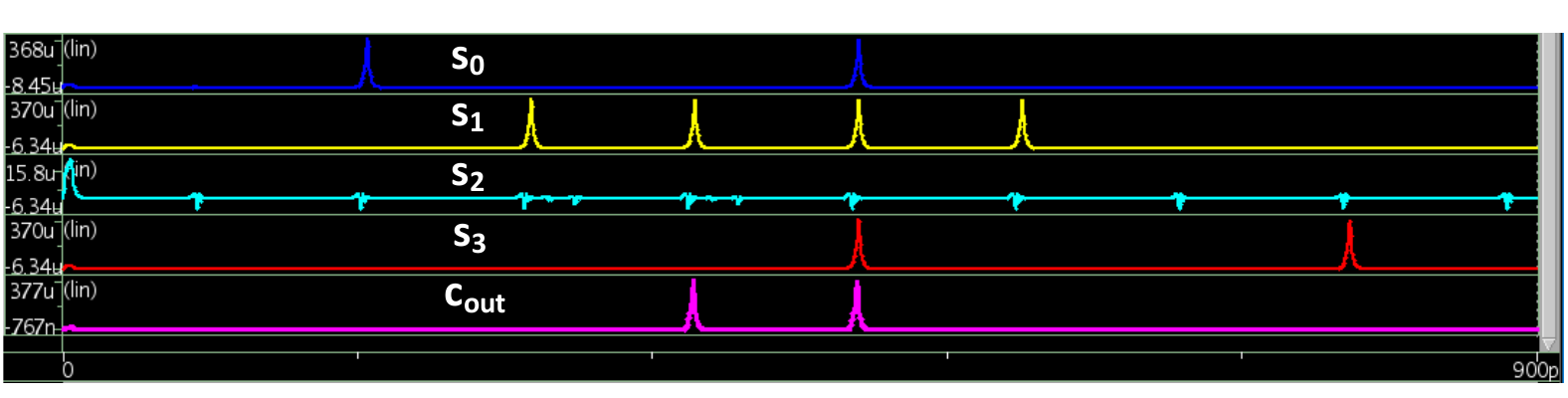}
                \caption{}
                \label{sim_2}
        \end{subfigure}
        \caption{Simulation results for a 4-bit Kogge-Stone adder (KSA4). (a) input/output signals for the KSA4 circuit generated by our algorithm, (b) output signals generated for the same inputs by a KSA4 circuit which is not path balanced. Four sets of random inputs are applied: $a_0$=1001, $a_1$=1111, $a_2$=1001, $a_3$=1111, $b_0$=1010, $b_1$=1100, $b_2$=1010, $b_3$=1100, $c_{in}$=1001. The correct outputs are: $S_0$=1010, $S_1$=1010, $S_2$=1110, $S_3$=1010, $C_{out}$=1101. As seen, only the results in (a) are correct.}
        \label{sim_example}
\end{figure*}
\section{Experimental Results}
\label{exper:sec}
The path balancing technology mapping algorithm (PBMap) is implemented inside the ABC \cite{synthesis2011abc}. Algorithm \ref{PBMap_alg} shows its pseudo code. After finding balanced matches for each node, depth of the best match is minimized using an algorithm similar to what is presented in \cite{cong1994flowmap}. This depth minimization is done without degrading the best achieved path balancing solution (without increasing the balancing overhead, i.e., \#DFFs). Faster system operation in the sense of finishing a given task in a shorter amount of time directly depends on the logical depth. In fact, the operation latency is the product of the number of cycles needed to do the operation and the clock cycle time. Shorter logical depth directly translates to lower cycle latency for the operation, but its effect on clock cycle time is hard to characterize before place and route is done. That is why we only talked about reducing the logical depth as our objective. 

One could consider the total number of gates and DFFs as the cost function and develop theorems similar to what is presented in Section \ref{DP_Tech-Map} for the new cost function. However, to minimize the total area, we added an extra area optimization pass as in line 9 of the shown pseudo code. In this area optimization pass, a match with the least area which preserves the best obtained \#DFFs and the minimum depth is chosen for each node.  
\begin{table*}[t]
\scriptsize
  \centering
  \caption{Experimental results for PBMap and baseline mapper (ABC's mapper). \#DFFs is reported for before and after applying the retiming algorithm. Area is in $mm^2$ and run-time is in $second$. Area and JJ count (\#JJs) are for after retiming. Logical depth is the maximum logic level in the network.}
    \begin{tabular}{ccccccccccccc}
    \toprule
     & \multicolumn{2}{c}{\#DFFs (before)} & \multicolumn{2}{c}{\#DFFs (after)}& \multicolumn{2}{c}{Area} & \multicolumn{2}{c}{\#JJ} & \multicolumn{2}{c}{Logical Depth}& \multicolumn{2}{c}{Run-time}\\
   circuits  &  PBMap & Baseline & PBMap & Baseline & PBMap & Baseline & PBMap & Baseline & PBMap & Baseline & PBMap & Baseline \\
    \toprule
c1908	&1033	&1216	&696	&844	&8.7	&9.3	&12013	&12785	&20	&24	&0.14	&0.21 \\
\midrule
c5315	&5289	&6146	&2908	&3575	&37.2	&42.1	&52033	&58661	 &23	&28 	&1.4	&2.1 \\
\midrule
c7552	&3681	&4354	&2429	&2867	&34.3	&37.4	&48482	&52641	  &19	 &22	 &1.04	 &1.9 \\
\midrule
c3540	&2683	&3187	&1159	&1372	&20.3	&21.8	&28300	&30165	&31	  &37	 &0.56	 &0.73 \\
\midrule
c499	&674	&632	&476	&444	&5.6	&5.6	&7758	&7734	 &13	 &13	&0.064  	&0.066  \\
\midrule
c880	&1406	&1663	&774	&957	&9.3	&10.4	&12909	&14415	&22 	&26 	&0.16	 &0.22  \\
\midrule
s1196	&1226	&1328	&746	&817	&11	&11.8	&15332	&16443	&18	    &20	   &0.29	  &0.25  \\
\midrule
s38417	&15929	&21289	&8405	&12306	&143	&168.7	&208289	&243091	  &21	&30	  &10.72	&17.13  \\
\midrule
s1238	&1558	&1665	&864	&984	&12.6	&13.8	&17617	&19171	  &19	&23	  &0.26  	&0.32  \\
\midrule
int2float	&507	&528	&270	&274	&4.5	&4.8	&6432	&6725	&16	 &16	&0.082	  &0.04  \\
\midrule
cavlc	&1514	&1544	&522	&565	&11.6	&12.2	&16339	&17115	&17	  &17	&0.19	  &0.02  \\
\midrule
priority	&9313	&35040	&9064	&19925	&71.9	&152.3	&102085	&212467	  &127	&249	&41.9	 &363.2  \\
\midrule
decoder 	&51	&51	&8	&8	&4	&6.2	&5469	&8340	&4	  &5	  &0.012	  &0.012   \\
\midrule
sin	  &75861	&89481	&13666	&16858	&153.8	&176.9	&215318	&245736	  &182	  &229	  &409.8	  &589.5  \\
\midrule
i10	  &11212	&15007	&7776	&10182	 &81.5	  &99.7	  &114306	&139263	  &33	  &43	  &5.6	   &10.7    \\
\midrule
frg2	&2796	&2974	&1375	&1470	&21.7	  &23.9	   &30340	  &33`237	  &12	  &13	  &0.53	   &0.62   \\
\midrule
pj1	   &66490	&83007	&36897	&43631	&411.1	   &468.2	 &585751	&663755	&34  	&44	   &115.8	   &186.6   \\
\midrule
i9	  & 1275	&1612	&647	&876	&12.8	   &14.9	&17842	    &20734	 &12	&15	   &0.26	    &0.3    \\
\midrule
9sym	&327	&353	&143	&149	&3.4	&3.6	&4859	  &5041	   &14	    &14   	&0.05	    &0.05    \\
\midrule
KSA4	& 30	&30     &25     	&25	        &0.5     	&0.5	      &692	  &692	  &6	   &6	    &0.02	   &0.0175\\
\midrule
KSA16	 &233	&235	&199	    &200	&3.4	    &3.5	     &4797	     &4842	      &10	     &10	&0.08	&0.07    \\
\midrule
ID8	    &4505	&5494	&1854	    &2140	 &16.1	    &19.4	    &22752	     &27020	   &77	   &85	     &2.32	   &3.32   \\

\bottomrule
avg. imp. & 20.64\% $\downarrow$& & 15.06\% $\downarrow$& & 12.22\% $\downarrow$& & 11.22\% $\downarrow$& & 14.56\% $\downarrow$& & 49.78\% $\downarrow$ \\

\bottomrule
    \end{tabular}%
  \label{exp_table}%
\end{table*}%

To see if circuits generated by PBMap operate correctly, we simulated a few benchmark circuits including a 4-bit Kogge Stone Adder (KSA4) using JSIM \cite{JSIM}. Fig. \ref{sim_example} shows the input/output waveform for the KSA4 circuit for two cases: (a) when our path balancing algorithm is applied, (b) without any path balancing. Four random values are considered for input $a$, input $b$, and carry in ($c_{in}$): $a_0$=1001, $a_1$=1111, $a_2$=1001, $a_3$=1111, $b_0$=1010, $b_1$=1100, $b_2$=1010, $b_3$=1100, $c_{in}$=1001. The correct sum ($S_0-S_{3}$) and carry out ($C_{out}$) for these inputs are as follows: $S_0$=1010, $S_1$=1010, $S_2$=1110, $S_3$=1010, $C_{out}$=1101. Please note that having a 0 as subscript of a signal makes it the least significant bit and having a 3 makes it the most significant bit. As seen in Fig. \ref{sim_1}, the circuit generated by PBMap produces the correct outputs, while as Fig. \ref{sim_2} shows, a circuit with no added path balancing DFFs produces erroneous outputs. Notice that since the depth of KSA4 circuit is 6, we have to wait at least 6 clock periods after applying the first set of inputs to see the first round of correct outputs.  

An SFQ library of gates as in \cite{RSFQLib}, consisting of \textit{and2}, \textit{or2}, \textit{xor2}, \textit{DFF}, \textit{splitter}, and \textit{inverter} gates were used, and several ISCAS \cite{hansen1999unveiling}, EPFL \cite{EPFL_bench}, MCNC \cite{yang1991logic}, and arithmetic benchmark circuits were considered. Table \ref{exp_table} show the experimental results for PBMap and a baseline mapper. The baseline mapper is ABC's mapper plus inserting path balancing DFFs and applying the standard retiming algorithm \cite{leiserson1991retiming} for minimizing DFF count. The total number of path balancing DFFs are mentioned for both before and after applying the retiming algorithm mainly to show the effectiveness of our path balancing algorithm in reducing DFF count. Retiming algorithm helps in reducing the DFF count in netlists generated by both PBMap and baseline mappers. In Table \ref{exp_table}, \textit{KSA16} and \textit{ID8} are 16-bit Kogge-Stone adder and 8-bit Integer Divider, respectively. 

PBMap was able to reduce \#DFFs by $2.7 \times$, and $1.2 \times$ before and after retiming for one of the EPFL benchmark circuits (\textit{priority}) compared to the baseline. PBMap reduces area, total JJ count (\#JJ), logical depth, and run-time by $1.11 \times$, $1.08 \times$, 96\%, and $7.66 \times$, respectively over the baseline for the same circuit. \textit{Area} in Table \ref{exp_table} is the total area of gates, path balancing DFFs, and splitters. On average for all benchmark circuits, PBMap improves the run-time over the baseline by 49.78\% mainly because its run-time for retiming is less than the baseline due to requirement of inserting fewer path balancing DFFs. \#DFFs (before retiming), \#DFFs (after retiming), area, \#JJs, and logical depth are reduced by an average of 20.64\%, 15.06\%, 12.22\%, 11.22\%, 14.56\%, respectively for PBMap compared with the baseline.

To compare circuits generated by PBMap with other published papers, we include experimental results of a 16-bit wave-pipelined sparse-tree RSFQ adder \cite{dorojevets201316}. The fabrication results published in \cite{dorojevets201316} shows that JJ count for this design is 9941. Using the same cell library (CONNECT cell library \cite{yorozu2002single}), PBMap consumes 8901 JJs for mapping a 16-bit Kogge-Stone adder which shows 10.5\% reduction in JJ count eventhough the Kogge-Stone adder is itself more complex than the sparse-tree adder. The difference between these two sets of numbers are because of the following reasons: (i) our algorithm is highly effective in reducing total JJ count. (ii) the results we presented are for post-synthesis i.e., we did not account for any JJs used in JTL connections, while the results in \cite{dorojevets201316} account for such connection costs. Please note that using CONNECT cell library, JJ count for KSA16 is increased compared with the case of using the cell library in \cite{RSFQLib}.  One important reason for seeing this difference is that there is no OR gate in the CONNECT cell library, and since inverter is expensive (it consumes 10 JJs while an XOR has 11 JJs and an AND gate has 13 JJs), implementing OR gate, which frequently appears in arithmetic circuits, using AND and inverter gates (De Morgan's law) consumes more JJs by a factor of 3 times or more. In \cite{filippov20118}, an 8-bit RSFQ ALU is presented, which supports 12 sets of operations including: \textit{ADD}, \textit{ADD-Invert A}, \text{AND}, \textit{NOR}, \textit{XNOR}. Therefore, this ALU is much more complex than our adders and comparing its JJ count with our adders is not fair, hence, it is not mentioned in this section.
\section{Conclusion}
\label{conc:sec}
In this paper, a path balancing technology mapping algorithm (PBMap) based on the dynamic rogramming approach is presented. We proved that the proposed technology mapping algorithm provides optimal path balancing tree mapping solutions for dc-biased SFQ circuits. Our technology mapping algorithm is quite effective in reducing the total number of path balancing DFFs for many tested benchmark circuits. To further reduce the total number of required path balancing DFFs, the standard retiming algorithm is used. Experimental results show that PBMap reduces \#DFFs, area, \#JJs, logical depth, and run-time by an average of 20.64\%, 12.22\%, 11.22\%, 14.56\%, and 49.78\%, respectively compared with the state-of-the-art technology mappers for 22 benchmark circuits. 
\section{Acknowledgement}
The research is based upon work supported by the Office of the Director of National Intelligence (ODNI), Intelligence Advanced Research Projects Activity (IARPA), via the U.S. Army Research Office grant W911NF-17-1-0120. The U.S. Government is authorized to reproduce and distribute reprints for Governmental purposes notwithstanding any copyright notation herein.
\ifCLASSOPTIONcaptionsoff
  \newpage
\fi
\bibliographystyle{IEEEtran}
\bibliography{IEEEabrv,jrnl}

\begin{IEEEbiography}[{\includegraphics[width=1.1 in, height=3.3 in, clip, keepaspectratio]{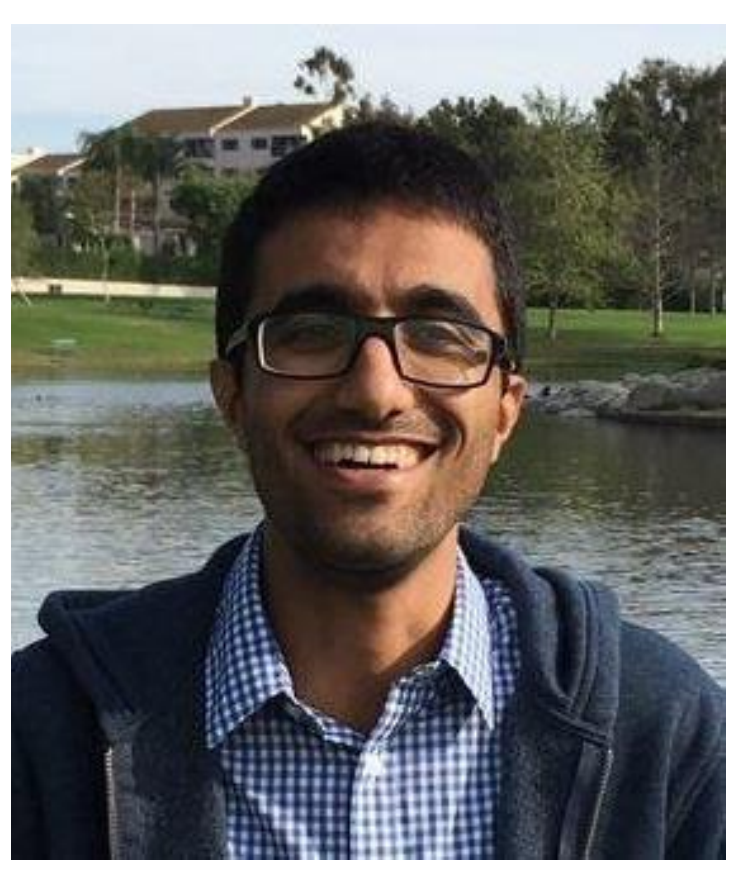}}] {Ghasem Pasandi}
(S'13) received the B.Sc, and M.Sc degrees in electrical and electronics engineering from the University of Tehran, Tehran, Iran in 2011, and 2014, respectively. He is currently a PhD student with the Ming Hsieh Department of Electrical Engineering, University of Southern California (USC), LA, CA. His research interests  include Computer Aided Design of Digital Systems, Superconductive Digital Electronics, Artificial Intelligence, Deep Learning, VLSI Implementation of Digital Systems, Low Power and Ultra Low Power Static Random Access Memory (SRAM) design, Low Power and Energy Efficient logic design.
\end{IEEEbiography}

\begin{IEEEbiography}[{\includegraphics[width=1.1 in, height=3.9 in, clip, keepaspectratio]{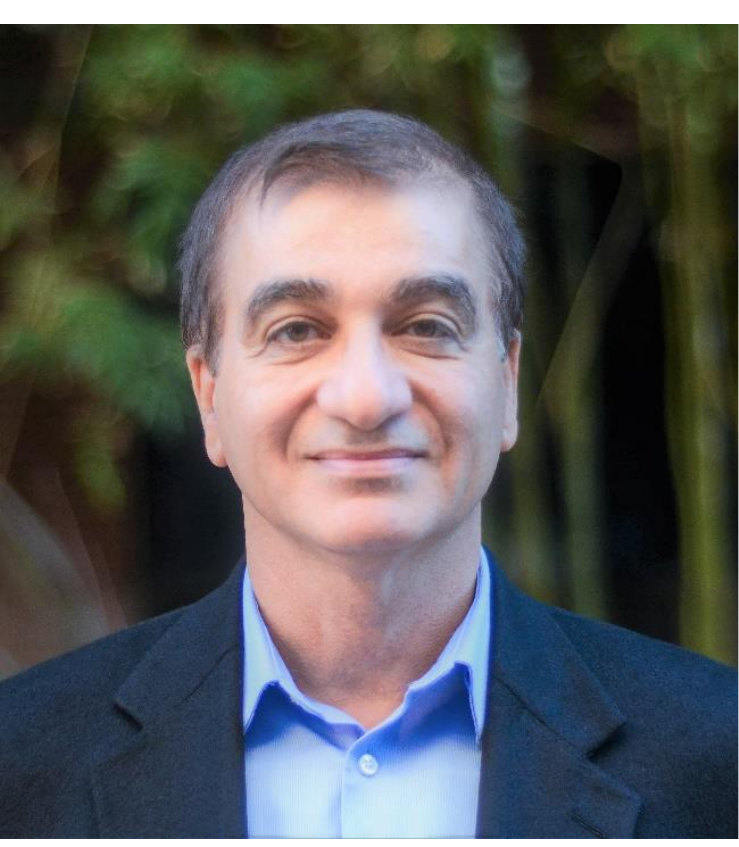}}]  {Massoud Pedram}
(F'01) obtained his B.S. degree in Electrical Engineering from the California Institute of Technology in 1986. Subsequently, he received M.S. and Ph.D. in Electrical Engineering and Computer Sciences from the University of California, Berkeley in 1989 and 1991, respectively. In September 1991, he joined the Ming Hsieh Department of Electrical Engineering of the University of Southern California where he currently is the Charles Lee Powell Professor of Electrical Engineering and Computer Science in the USC Viterbi School of Engineering. Dr. Pedram is a recipient of the IEEE Circuits and Systems Society Charles A. Desoer Technical Achievement Award (2015), the Presidential Early Career Award for Scientists and Engineers (1996), and the National Science Foundation's Young Investigator Award (1994). His research has received a number of other awards including two Design Automation Conference Best Paper Awards, a Distinguished Paper Citation from the Int'l Conference on Computer Aided Design, one Best Paper Award of the ACM/IEEE Int’l Symp. on Low Power Design and Electronics, three Best Paper Awards from the International Conference on Computer Design, one Best Paper Award of the IEEE Computer Society Annual Symp. on VLSI, an IEEE Transactions on VLSI Systems Best Paper Award, and an IEEE Circuits and Systems Society Guillemin-Cauer Award. Dr. Pedram was recognized as one of the four DAC Prolific Authors (with 50+ papers) and the DAC Bronze Cited Author at the 50th anniversary of the Design Automation Conf., Austin, TX (2013), received a Frequent Author Award (Top Three Author Award) at the 20th Anniversary Asia and South Pacific Design Automation Conference, Chiba/Tokyo, Japan (2015), and listed as the Second Most Prolific and Second Most Cited Author at the 20th Anniversary Int'l Symp. on Low Power Electronics and Design, Rome, Italy (2015).
\end{IEEEbiography}
\appendices

\section{Proof of lemma 1}
\label{appendix_a}
We use induction hypothesis for proving lemma 1. \textit{Base case}: a tree with one node has two input pins. \textit{Induction step}: assume that for a binary tree with $N$ internal nodes, there are $N$+$1$ input pins. Now, for the $N$+$1$ step, we need to add one more node to the previous tree by replacing an input pin with a new node. If an input pin is replaced by a new node, both the number of nodes and the number of input pins increase by one (one input pin is lost and two new input pins are gained in return for the new added node.)$\blacksquare$
\begin{figure}[b]
\centering
\includegraphics[width=0.47\textwidth]{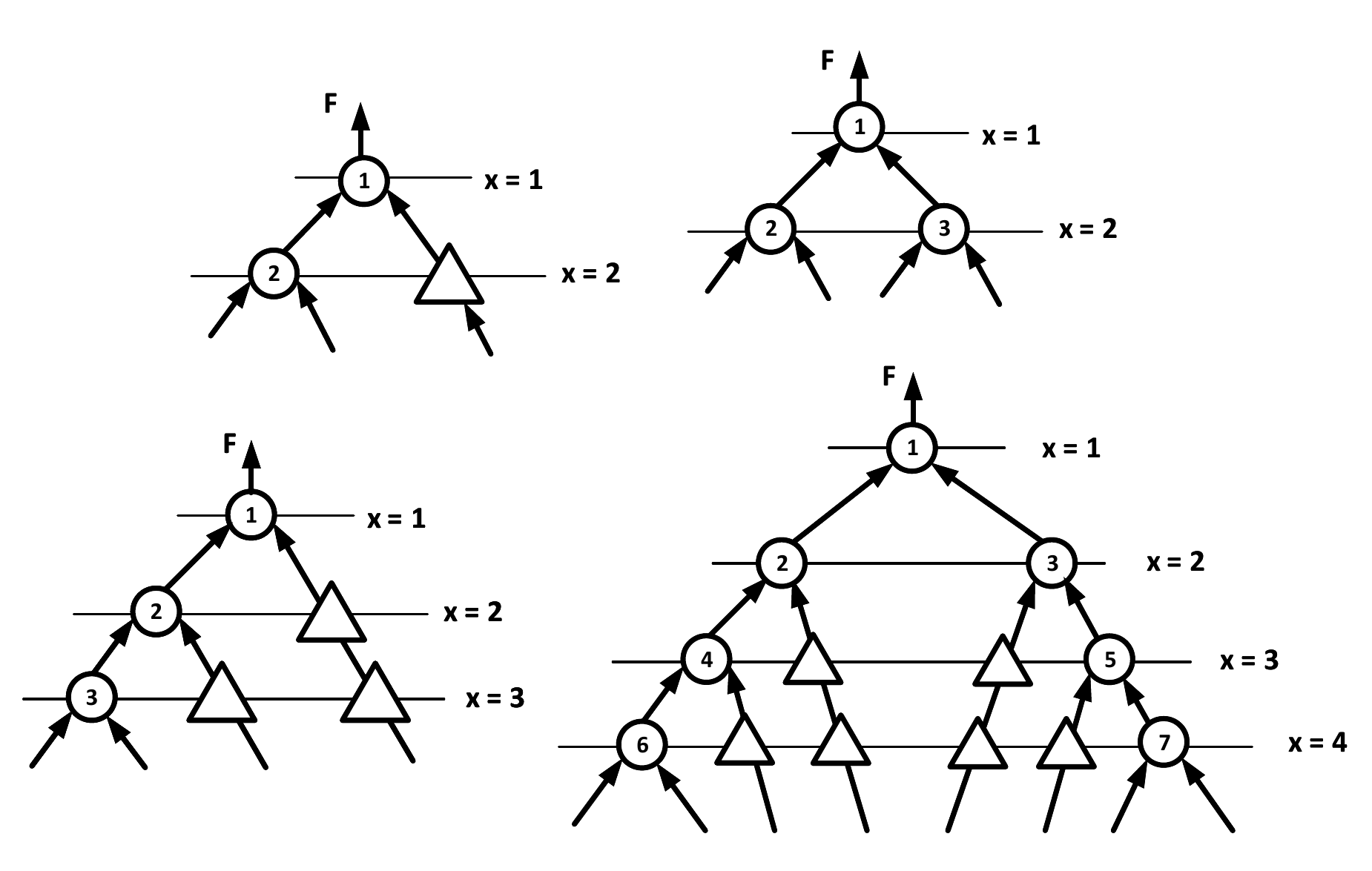}
\caption{Two possible binary trees with height $X=2$, and the most unbalanced binary trees with height $X$=$3$, and $X$=$4$.}
\label{X_2_3}
\end{figure}
\begin{figure*}[t]
\centering
\includegraphics[width=0.75\textwidth]{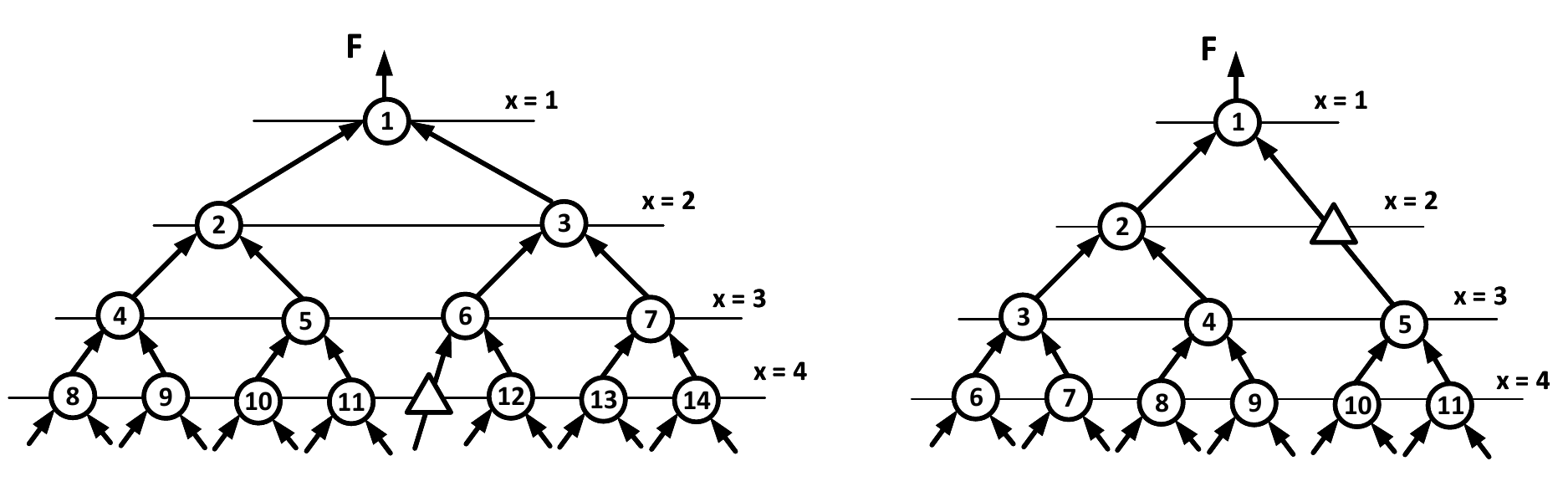}
\caption{Two examples for lemma 5.}
\label{X=4}
\end{figure*}
\section{Proof of lemma 2}
\label{appendix_b}
Since \#input pins is $n$, based on lemma 1, the total number of nodes is fixed at $n$-$1$. Now, if we want to create a tree with the maximum height (to maximize $p$), a single node should be put at each level, because at least one node has to be present at each level. Thus, the maximum height will be $n$-$1$, hence, $X$+$p \leq n$-$1$. Therefore, $p \leq n$-$1$-$X$. $\blacksquare$
\section{Proof of lemma 3}
\label{appendix_c}
Since $X$ is not a large number, it is easy to manually check the correctness of this lemma. Fig. \ref{X_2_3} shows two possible binary trees for $X$=$2$, and the most unbalanced binary tree for $X=3$. The buffer nodes are shown in these figures too. In the most unbalanced case there is one buffer node at level two, two buffer nodes at level three, three buffer nodes at level four ,..., $X$-$1$ buffer nodes at the last level. So, the total number of buffer nodes is equal to the sum of the natural numbers from $1$ to $X$-$1$, which is $(X$-$1)X$/$2$. $\blacksquare$
\section{Proof of lemma 4}
\label{appendix_d}
The most unbalanced binary tree with height $X\geq 4$ is achieved when we start placing a node at each level from level $x$=$X$ to the level $x$=$1$ (until now, $X$ nodes are consumed), and returning from the right side of the tree with similar method, consuming $X$-$1$ more nodes. So, the total number of nodes will be $2X$-$1$, and based on lemma 1, \#input pins is $2X$. From now on, adding more nodes means removing one buffer node, so, making the tree more balanced. The total number of buffer nodes is computed similar to lemma 3.$\blacksquare$
\section{Proof of lemma 5}
\label{appendix_e}
To have the most balanced binary tree with height $X$ and \#input pin of $n$, we need to find a tree with minimum total buffer node count, or equivalently the minimum value for $Y$ (sum of the $y_i$s). 
By fixing the values of $H$ and $n$ in Eq(\ref{n_formula3}), the minimum value for $Y$ is obtained when a $y_i$ with a larger coefficient contributes more in this equation. In other words, we should start with $y_2$ and maximize it, then if more values are needed to satisfy Eq(\ref{n_formula3}), $y_3$ should be maximized. We should continue this way until the left hand side of Eq(\ref{n_formula3}) becomes equal to its right hand side. If this happens at level $i$, then $y_j$=$0$ for all $j > i$. It may also be needed to add a few buffer nodes to level $i$+$1$ to satisfy the Eq(\ref{n_formula3}) without maximizing $y_{i+1}$.$\blacksquare$

Fig. \ref{X=4} shows two examples for $X$=$4$. As seen in the left tree with $n$=$15$, the maximum value for $y_2$ cannot be used because it does not generate a valid tree based on the given constraints for height and \#input pins. Thus, $y_2$ is set to 0. For this tree, the only scenario which satisfies Eq(\ref{n_formula3}) is when there is a buffer node at the last level, as shown in this figure. However, for $n$=$12$, we can set $y_2$=$1$, which generates the right graph in Fig. \ref{X=4}.
\section{Proof of lemma 6}
\label{appendix_f}
\begin{figure}[t]
\centering
\includegraphics[width=0.45\textwidth]{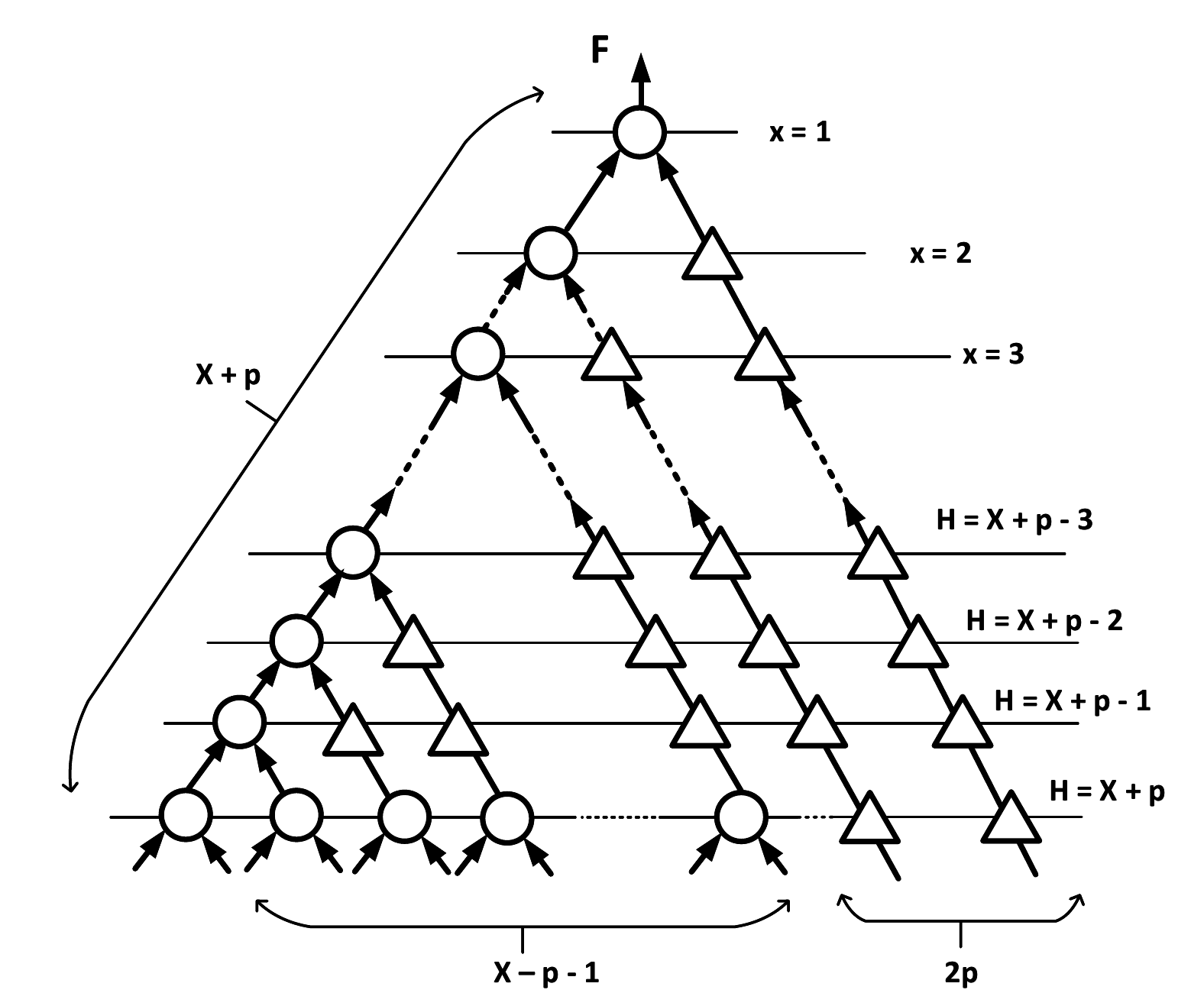}
\caption{The most balanced binary tree that we can get by increasing the height of the tree described in lemma 4 without increasing the number of input pins.}
\label{Most_Balance}
\end{figure}
Based on lemma 5, to find the most balanced binary tree, we should start consuming buffer nodes at the lower levels (closer to the root of the tree). In the case of a tree described in lemma 4, there are $2X$-$1$ nodes. $X$+$p$ nodes should be put at each level (one per level) to generate the height of $X$+$p$. The rest of the nodes, $X$-$p$-$1$, are put at the last level of the tree. So, a tree similar to what is shown in Fig. \ref{Most_Balance} will be obtained. Now, we need to count the number of buffer nodes of this tree. It consists of two groups. First, the $X$-$p$-$1$ nodes at the last level. The first of these nodes needs $0$ buffer node, the second one needs $1$, the third one $2$,..., the $(X$-$p$-$1)^{th}$ one needs $X$-$p$-$2$ buffer nodes. This sums up to $(X$-$p$-$2)(X$-$p$-$1)/2$. The second group of buffer nodes correspond to the long wires which start from a node at level 1 to level $2p$ and end at the last level of the tree. The first wire in this group which starts from the node at level $x$=$1$, needs $X$-$p$-$1$ buffer nodes because it travels from level $2$ to the last level of the tree and at each level it needs one buffer node. The second wire in this group needs $X$-$p$-$2$ buffer nodes,..., summing up to $2pX$+$p$-$2p^2$. Therefore, the total number of buffer nodes are $(X$-$p$-$1)(X$-$p$-$2)/2$+$2pX$+$p$-$2p^2$. $\blacksquare$
\section{Proof of lemma 7}
\label{appendix_g}
Referring to Section \ref{proof:sec}, the formula of \#input pins for a binary tree is developed by considering the effect of each buffer node at each level of the tree in reducing the total number of input pins compared with a full binary tree. After retiming, a subset of buffer-nodes will be moved from higher levels (closer to leaves of the tree) to lower levels (closer to the root of the tree). Therefore, it should be proven that the contribution of buffer-nodes in Eq(\ref{n_formula2}) for new architecture (after retiming) and for the old one (before retiming) are the same. In other words, Eq(\ref{n_formula2}) is valid for before and after retiming. For this purpose, suppose that there is a node (node $j$) at level $x$=$X$ of a binary tree and for simplicity, suppose that all of its inputs are coming from PIs (e.g. node $3$ in Fig. \ref{ret1}). This node generates two buffer nodes per level starting from $x$=$X$+$1$ all the way down to the last level ($x$=$H$). The contribution of those buffer nodes in the right-hand side of Eq(\ref{n_formula2}) is as follows:
\begin{equation}
\label{eq_ret1}
2\times \lbrace 2^{H-(X+1)} + 2^{H-(X+2)} + ... + 2^1 + 1 \rbrace
\end{equation} 

After retiming, node $j$ will be moved to the last level ($x$=$H$), and it will generate a single buffer node at each level starting from $x$=$H$-$1$ all the way up to $x$=$X$. For example, after retiming is applied to the tree in Fig. \ref{ret1}, node $3$ will be moved to the last level ($x$=$4$) and it will generate one buffer node at level three and one at level two. Contributions of the new buffer nodes generated after retiming to the right-hand side of Eq(\ref{n_formula2}) is as follows:
\begin{equation}
\label{eq_ret2}
1\times \lbrace 2^{H-(X)} + 2^{H-(X+1)} + ... + 2^1 \rbrace
\end{equation} 

It is easy to see that both of Eq(\ref{eq_ret1}) and Eq(\ref{eq_ret2}) have the same values equal to $2^{H-X+1} -2$. This shows that Eq(\ref{n_formula2}) is valid for before and after retiming, hence, lemma 7 is proven.$\blacksquare$

\end{document}